\documentclass[12pt,a4paper]{article}

\usepackage{graphics}
\usepackage{amsmath}
\usepackage{amssymb}

\title{Higgs production at a muon collider in the Two Higgs Doublet Model 
of type II}
\author{Carlos A. Mar\'{\i}n}
\date{\small{Universidad San Francisco de Quito \\
29 April 2004}}
\begin{document}
\maketitle

\begin{abstract}
\noindent
We calculate Higgs production cross sections at a muon collider
in the Two Higgs Doublet Model of type II. The most interesting
productions channels are $\mu^- \mu^+ \rightarrow h^0 Z^0, H^0
Z^0, H^- H^+, A^0Z^0$ and $H^{\mp} W^{\pm}$. The last channel is 
compared with the production processes $p \bar{p} \rightarrow
H^{\mp} W^{\pm} X$ and $p p \rightarrow H^{\mp} W^{\pm} X$ at the 
Tevatron and LHC energies, respectively, for large values of 
$\tan\beta$.
\end{abstract}

\tableofcontents

\section{Introduction}
In this article we calculate neutral and charged Higgs production
cross sections at a muon collider in the Two Higgs Doublet Model
of type II. The Higgs sector of the Minimal Supersymmetric
Standard Model (MSSM) is of 
this type (tho the model of type II does not require Supersymmetry).
Higgs doublets can be added to the Standard Model without upsetting the
$Z/W$ mass ratio. Higher dimensional representations upset this ratio
\cite{B-H}. Adding a second complex doublet to the Standard Model results 
in five Higgs bosons: one pseudoscalar $A^0$ (CP-odd scalar), two neutral
scalars $H^0$ and $h^0$ (CP-even scalars), and two charged scalars $H^+$ 
and $H^-$. In the Standard Model we only have a single neutral Higgs.

In recent years, some papers have appeared,
suggesting the possibility of the 
construction of a $\mu^- \mu^+$ collider to detect charged or neutral
Higgs bosons [\cite{mu1}, \cite{mu2}]. The main reason is that in
a muon collider, the signal would be much cleaner than in a hadron
collider. In this paper, we analyze this possibility studying some 
production cross sections like:
 $\mu^- \mu^+ \rightarrow h^0 Z^0, H^0   
Z^0, H^- H^+, A^0Z^0$ and $H^{\mp} W^{\pm}$ (Sections 2-6). 

In Sections 5,6,8,9 we will focus our interest in the production of 
charged Higgs bosons. There are three ways of producing $H^{\pm}$.
One is via $p \bar{p}$ or $pp$ interactions in a hadron collider. In 
hadron colliders, the signals are overwhelmed by  backgrounds due 
basically to $t\bar{t}$ production \cite{S.Moretti}. The other ways 
to produce charged Higgs are  $e^- e^+$ or $\mu^- \mu^+$ colliders 
, in which backgrounds are considerably less. In some 
processes like $\mu^- \mu^+ \rightarrow H^- H^+$ and $e^- e^+ \rightarrow
H^- H^+$, there is no difference between the cross sections obtained in an 
$e^- e^+$ collider or a $\mu^- \mu^+$ collider. However, in reactions like
$\mu^- \mu^+ \rightarrow H^{\mp} W^{\pm}$ and $e^- e^+ \rightarrow H^{\mp} 
W^{\pm}$, the total cross section is proportional to the square of the 
mass of the fermion and then $e^- e^+$ interactions give us very small 
cross sections. This motivated us to compare in Section
9  the channel $\mu^- \mu^+ \rightarrow H^{\mp} W^{\pm}$
(at $\sqrt{s} = 500\textrm{GeV/c}$ and for large values of $\tan\beta$) 
with the production processes  $p 
\bar{p} 
\rightarrow H^{\mp} W^{\pm} X$ (at the Tevatron) and $p p 
\rightarrow H^{\mp} W^{\pm} X$ (at the LHC), to check the feasibility
of detecting $H^{\pm}$ using a muon collider.

The influence of radiative corrections in the masses of the Higgs 
bosons is  considered in all the calculations. 

\section{Higgs bosons masses and radiative corrections}
The masses of the neutral Higgs particles, calculated at tree level,
are \cite{V.Barger}:

\begin{equation}
m_{A^0}^2 = m_H^2 - m_W^2
\label{mAo}
\end{equation}

\begin{equation}
m_{H^0}^2 = \frac{1}{2} \left[ m_Z^2 + m_{A^0}^2
+ \left[ \left(m_Z^2 - m_{A^0}^2 \right)^2 +
4 m_{A^0}^2 m_Z^2 \sin^2 2\beta \right]^{1/2}
\right]
\label{mHo}
\end{equation}

\begin{equation}
m_{h^0}^2 = \frac{1}{2} \left[ m_Z^2 + m_{A^0}^2
- \left[ \left(m_Z^2 - m_{A^0}^2 \right)^2 +
4 m_{A^0}^2 m_Z^2 \sin^2 2\beta \right]^{1/2}
\right]
\label{mhowrc}
\end{equation}

\noindent with $ 0 \le \beta < \frac{\pi}{2}$

From these relations, the Higgs bosons masses satisfy the
bounds:

\begin{equation}
m_{A^0} < m_H
\label{mA<mH}
\end{equation}

\begin{equation}
m_H > m_W
\label{mH>mW}
\end{equation}

\begin{equation}
m_{h^0} \le m_Z
\label{mh<mZ}
\end{equation}

\begin{equation}
m_Z \le m_{H^0} \le \sec\theta_W m_H
\label{mHbounds}
\end{equation}

The bound given by (\ref{mh<mZ}) practically has been
excluded by the present limits on $m_{h^0}$ obtained by LEP
and CDF \cite{LEP_CDF}.

The mixing angle $\alpha$ $( - \pi/2 < \alpha \le 0)$
between the two neutral scalar Higgs
fields $H^0$, $h^0$ is given by

\begin{equation}
\tan\alpha = -\left[ \frac{1 + F}{1 - F} \right]^{1/2}
\label{tan_alpha}
\end{equation}
\begin{equation}
F = \frac{\left( 1 - \tan^2\beta \right)}
{ \left( 1 + \tan^2 \beta \right) G}
\left[ 1 - \frac{m_Z^2}{m_H^2} - \frac{m_W^2}{m_H^2} \right]
\label{F}
\end{equation}

\begin{equation}
G = \left[ \left( 1 - \frac{m_W^2}{m_H^2}
+ \frac{m_Z^2}{m_H^2} \right)^2 -
4 \left(\frac{m_Z^2}{m_H^2} \right)
\left( 1 - \frac{m_W^2}{m_H^2} \right)
\left( \frac{1 - \tan^2\beta}{1 + \tan^2\beta}
\right)^2 \right]^{1/2}
\label{G*}  
\end{equation}

In terms of $m_H$ and $G$ Equations (\ref{mHo}) and
(\ref{mhowrc}) are:

\begin{equation}
m_{H^0}^2 = \frac{1}{2} m_H^2 \left[ 1 - \frac{m_W^2}{m_H^2} +
\frac{m_Z^2}{m_H^2} + G \right]
\label{mHoG}
\end{equation}

\begin{equation}
m_{h^0}^2 = \frac{1}{2} m_H^2 \left[ 1 - \frac{m_W^2}{m_H^2} +
\frac{m_Z^2}{m_H^2} - G \right]
\label{mhoG}
\end{equation}

Taking into account radiative corrections, (\ref{mHo})
and (\ref{mhowrc}) can be written as [see \cite{Weinberg},
\cite{Zhou}]:

\begin{eqnarray}
\lefteqn{
m_{H^0}^2 = \frac{1}{2} \{ m_A^2 + m_Z^2 + \Delta_t
+ \Delta_b
}
\nonumber \\ & &
+ \left[ \left( \left( m_A^2 - m_Z^2 \right) \cos 2\beta
+ \Delta_t - \Delta_b \right)^2 +
\left( m_A^2 + m_Z^2 \right)^2 \sin^2 2\beta \right]^{1/2}
\}
\label{radiative_mHo}
\end{eqnarray}

\begin{eqnarray}
\lefteqn{
m_{h^0}^2 = \frac{1}{2} \{ m_A^2 + m_Z^2 + \Delta_t
+ \Delta_b
}
\nonumber \\ & &
- \left[ \left( \left( m_A^2 - m_Z^2 \right) \cos 2\beta
+ \Delta_t - \Delta_b \right)^2 +
\left( m_A^2 + m_Z^2 \right)^2 \sin^2 2\beta \right]^{1/2}
\}
\label{radiative_mho}
\end{eqnarray}

\noindent where:
\begin{equation} 
\Delta_b = \frac{3 \sqrt{2} m_b^4 G_F \left( 1 +
\tan^2\beta \right)}{2 \pi^2 }
ln\left( \frac{M_{sb}^2}{m_b^2} \right)  
\label{Delta_b}
\end{equation}

\noindent and
\begin{equation}
\Delta_t = \frac{3 \sqrt{2} m_t^4 G_F \left( 1 +
\tan^2\beta \right)}{2 \pi^2 \tan^2\beta}
ln\left( \frac{M_{st}^2}{m_t^2} \right)
\label{Delta_t}
\end{equation}

\noindent $M_{sb}$ and $M_{st}$ are the masses of the sbottom and stop
(the scalar superpartners of the bottom and top quarks).

Equation (\ref{mAo}) is practically unaffected by radiative 
corrections. According to (\ref{radiative_mho})
 $m_{h^0}$ increases
as the value of $m_A$ increases. Then, for very large values of
$m_A$ we can set an upper bound for $m_{h^0}$:

\begin{equation}
m_{h^0} \le m_{h^0} (m_{A^0} \rightarrow \infty) =
m_Z^2 \left( \frac{1 - \tan^2\beta}{1 + \tan^2\beta} \right)^2 
+ \frac{ \Delta_t \tan^2\beta}{\left( 1 + \tan^2\beta \right)}
+ \frac{\Delta_b}{\left( 1 + \tan^2\beta \right)}
\label{mhupperlimit}
\end{equation} 

Taking $m_b = 4.3$ $\textrm{GeV/c}^2$, $m_t = 174.3 \textrm{GeV/c}^2$,
$M_{st} \sim M_{sb} \sim 1 \textrm{TeV}$ \cite{Weinberg} and $m_Z = 
91.1876 \textrm{GeV/c}^2$
we obtain:

\begin{equation}
\Delta_b = 1.123 \times 10^{-6} 
\left( 1 + \tan^2\beta \right) m_Z^2
\label{Db}
\end{equation}

\begin{equation}
\Delta_t = 0.9723 m_Z^2 \frac{\left(1 + \tan^2\beta \right)}
{\tan^2\beta}
\label{Dt}
\end{equation}

The contribution of the b-quark loop is negligible.
Using Equations (\ref{Db}) and (\ref{Dt}), 
(\ref{mhupperlimit}) can be expressed as:

\begin{equation}
m_{h^0} \le m_Z \left[ \left( \frac{ 1 - \tan^2\beta}
{1 + \tan^2\beta} \right)^2 + 0.9723 \right]^{1/2}
\label{mholimit}
\end{equation}

For large values of $\tan\beta$ ($\tan\beta \rightarrow
\infty$) we obtain the limit

\begin{equation}
m_{h^0} \le 1.4044 m_Z = 128.062 \textrm{GeV/c}^2
\label{mhofinallimit}
\end{equation}

The upper bound on $m_{h^0}$ is raised by radiative corrections from
$m_Z$ to 128.062 $\textrm{GeV/c}^2$ for stop masses of order 1 TeV.

Considering radiative corrections, we can write, for the masses of
the neutral Higgs scalars:

\begin{equation}
m_{H^0}^2 = \frac{1}{2} m_H^2 \left[ 1 - \frac{m_W^2}{m_H^2} +
\frac{m_Z^2}{m_H^2} +\frac{\Delta_t}{m_H^2} + G_{rc} \right]
\label{mHoGrc}
\end{equation}

\begin{equation}
m_{h^0}^2 = \frac{1}{2} m_H^2 \left[ 1 - \frac{m_W^2}{m_H^2} +
\frac{m_Z^2}{m_H^2} + \frac{\Delta_t}{m_H^2} - G_{rc} \right]
\label{mhoGrc}
\end{equation}

\begin{eqnarray}
\lefteqn{
G_{rc} = [ \left( 1 - \frac{m_W^2}{m_H^2}
+ \frac{m_Z^2}{m_H^2} \right)^2 -
4 \left(\frac{m_Z^2}{m_H^2} \right)
\left( 1 - \frac{m_W^2}{m_H^2} \right)   
\left( \frac{1 - \tan^2\beta}{1 + \tan^2\beta}
\right)^2
}
\nonumber \\ & &
+ 2 \left(\frac{\Delta_t}{m_H^2} \right) 
 \left( 1 - \frac{m_W^2}{m_H^2}
- \frac{m_Z^2}{m_H^2} \right)
\left( \frac{1 - \tan^2\beta}{1 + \tan^2\beta}
\right)
 + \left( \frac{\Delta_t}{m_H^2}
 \right)^2 ]^{1/2}
\label{Grc}
\end{eqnarray}

With radiative corrections, the value of the $\alpha$ parameter 
is:

\begin{equation}
\tan\alpha = -\left[ \frac{1 + F_{rc}}{1 - F_{rc}} \right]^{1/2}
\label{tan_alpharc}
\end{equation}

\begin{equation}
F_{rc} = \frac{\left[\left(\frac{ 1 - \tan^2\beta }
{ 1 + \tan^2 \beta }\right)
\left( 1 - \frac{m_Z^2}{m_H^2} - \frac{m_W^2}{m_H^2} \right)
+ \frac{\Delta_t}{m_H^2} \right]}{G_{rc}}
\label{Frc}
\end{equation}

Additionally we have:

\begin{equation}
\sin2\alpha = - \frac{2 \tan\beta}{\left(1 + \tan^2\beta
\right)} \frac{\left( 1 - \frac{m_W^2}{m_H^2} + \frac{m_Z^2}
{m_H^2} \right)}{G_{rc}}
\label{sin2alpha}
\end{equation}

\section{Production of $h^0$ , $H^0$}
\begin{figure}
\begin{center}
%\vspace*{-4.5cm}
%\scalebox{0.5}
{\includegraphics{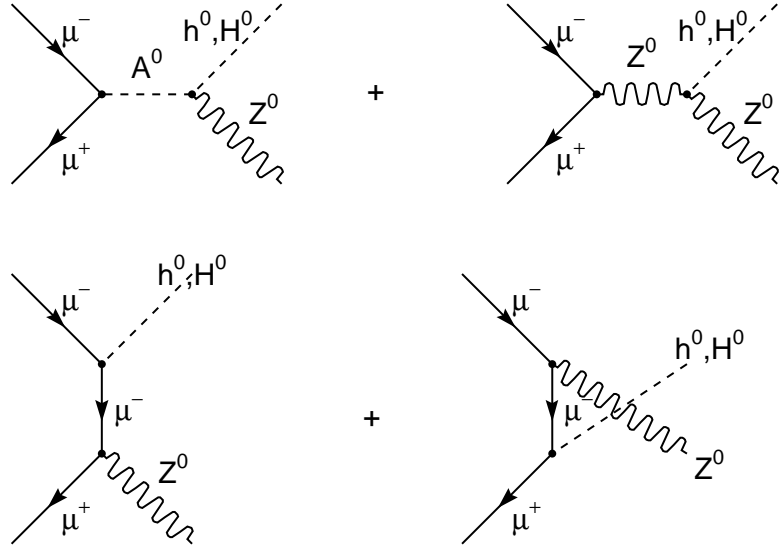}}
%\vspace*{0.7cm}
\caption{Feynman diagrams corresponding to the production
of $h^0$ or $H^0$ in the channel $\mu^- \mu^+ \rightarrow h^0 Z^0$.}
\label{mumu_hZ_fig}
\end{center}
\end{figure}

From the Feynman diagrams in Figure \ref{mumu_hZ_fig} and the
corresponding Feynman rules given in reference \cite{M_H}, we obtain
the differential cross section for the reaction  $\mu^- \mu^+ 
\rightarrow h^0 Z^0$ in the center of mass system

\begin{eqnarray}
\lefteqn{
\frac{d \sigma}{d \Omega} (\mu^- \mu^+ \rightarrow h^0 Z^0) =
\frac{1}{ 64 \pi^2 s^2} G_F^2 m_Z^4 \left| C_Z \right|^2 
\Lambda^{1/2} ( s, m_{h^0}^2, m_Z^2 )
}
\nonumber \\
& &
\left[  \left( g_A^\mu \right)^2 + \left( g_V^\mu \right)^2 
\right] \left[ 8 s m_Z^2 + \Lambda ( s, m_{h^0}^2, m_Z^2 ) 
sin^2 \theta \right]
\label{muon_hZ}
\end{eqnarray}
where 
\begin{eqnarray}
\lefteqn{
g_A^\mu = - \frac{1}{2}
}
\nonumber \\ & 
g_V^\mu = - \frac{1}{2} + 2 \sin^2 \theta_W
\label{gA,gV}
\end{eqnarray}

\begin{equation}
\Lambda (a, b, c) = a^2 + b^2 + c^2 - 2ab - 2ac - 2bc
\label{Lambda}
\end{equation}
\begin{equation}
 C_Z  = \frac{\sin \left( \beta - \alpha \right)}
{\left( s - m_Z^2 + i m_Z \Gamma_Z \right)}
\label{CZ}
\end{equation}
$\Gamma_Z$ is the total decay width of the $Z^0$ and $\theta$
is the scattering angle in the center of mass system. 

The total cross section corresponding to $\mu^- \mu^+ \rightarrow h^0 
Z^0$ is obtained integrating  Equation (\ref{muon_hZ}):

\begin{eqnarray}
\lefteqn{
\sigma (\mu^- \mu^+ \rightarrow h^0 Z^0) = \frac{G_F^2 m_Z^4 
\left( \tan\beta - \tan\alpha \right)^2}
{48 \pi s^2 \left( 1 + \tan^2\alpha \right) \left( 1 + 
\tan^2 \beta \right)}
}
\nonumber \\ & &
\times
\frac{\left( 1 - 4 \sin^2 \theta_W + 8 \sin^4 \theta_W \right)}
{\left[ \left( s - m_Z^2 \right)^2 + m_Z^2 \Gamma_Z^2 \right]}
\left[ 12 s m_Z^2 + \Lambda ( s, m_{h^0}^2, m_Z^2 ) \right]
\nonumber \\ & &
\times\Lambda^{1/2} ( s, m_{h^0}^2, m_Z^2 ) \times 
\left( 3.8938 \times 10^{11} \right) \textrm{fb}
\label{sigma_muhz}
\end{eqnarray}

\begin{figure}
\begin{center}
\input{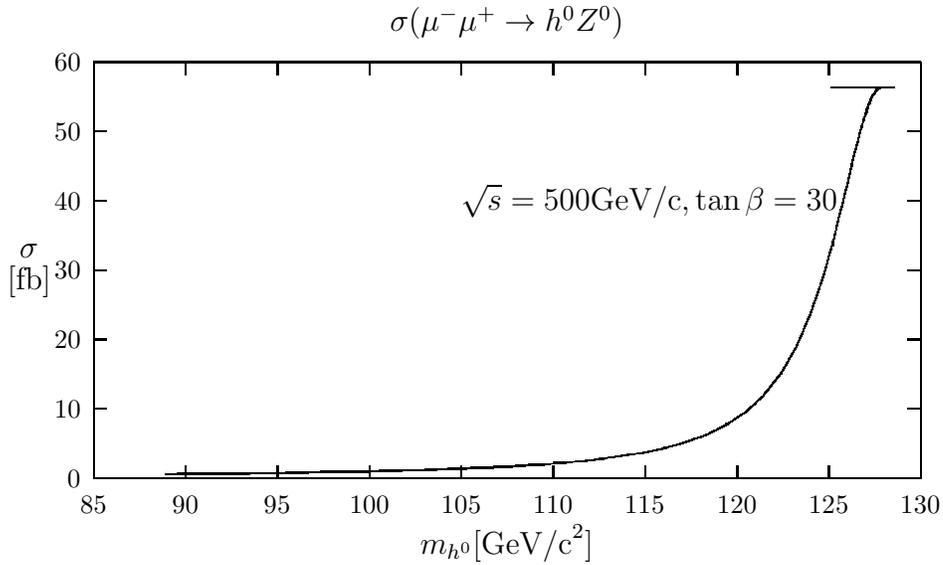}
\caption{Total cross section for the process
$\mu^- \mu^+ \rightarrow h^0 Z^0$ as a function
of $m_{h^0}$. We have taken
$\sqrt{s} = 500 \textrm{GeV/c}$ and $\tan\beta= 30$.}
\label{mumuhz_graph1}
\end{center}
\end{figure}

\begin{figure}
\begin{center}
% GNUPLOT: LaTeX picture
\setlength{\unitlength}{0.240900pt}
\ifx\plotpoint\undefined\newsavebox{\plotpoint}\fi
\sbox{\plotpoint}{\rule[-0.200pt]{0.400pt}{0.400pt}}%
\begin{picture}(1500,900)(0,0)
\font\gnuplot=cmr10 at 10pt
\gnuplot
\sbox{\plotpoint}{\rule[-0.200pt]{0.400pt}{0.400pt}}%
\put(161.0,123.0){\rule[-0.200pt]{4.818pt}{0.400pt}}
\put(141,123){\makebox(0,0)[r]{0}}
\put(1419.0,123.0){\rule[-0.200pt]{4.818pt}{0.400pt}}
\put(161.0,188.0){\rule[-0.200pt]{4.818pt}{0.400pt}}
\put(141,188){\makebox(0,0)[r]{10}}
\put(1419.0,188.0){\rule[-0.200pt]{4.818pt}{0.400pt}}
\put(161.0,254.0){\rule[-0.200pt]{4.818pt}{0.400pt}}
\put(141,254){\makebox(0,0)[r]{20}}
\put(1419.0,254.0){\rule[-0.200pt]{4.818pt}{0.400pt}}
\put(161.0,319.0){\rule[-0.200pt]{4.818pt}{0.400pt}}
\put(141,319){\makebox(0,0)[r]{30}}
\put(1419.0,319.0){\rule[-0.200pt]{4.818pt}{0.400pt}}
\put(161.0,385.0){\rule[-0.200pt]{4.818pt}{0.400pt}}
\put(141,385){\makebox(0,0)[r]{40}}
\put(1419.0,385.0){\rule[-0.200pt]{4.818pt}{0.400pt}}
\put(161.0,450.0){\rule[-0.200pt]{4.818pt}{0.400pt}}
\put(141,450){\makebox(0,0)[r]{50}}
\put(1419.0,450.0){\rule[-0.200pt]{4.818pt}{0.400pt}}
\put(161.0,515.0){\rule[-0.200pt]{4.818pt}{0.400pt}}
\put(141,515){\makebox(0,0)[r]{60}}
\put(1419.0,515.0){\rule[-0.200pt]{4.818pt}{0.400pt}}
\put(161.0,581.0){\rule[-0.200pt]{4.818pt}{0.400pt}}
\put(141,581){\makebox(0,0)[r]{70}}
\put(1419.0,581.0){\rule[-0.200pt]{4.818pt}{0.400pt}}
\put(161.0,646.0){\rule[-0.200pt]{4.818pt}{0.400pt}}
\put(141,646){\makebox(0,0)[r]{80}}
\put(1419.0,646.0){\rule[-0.200pt]{4.818pt}{0.400pt}}
\put(161.0,712.0){\rule[-0.200pt]{4.818pt}{0.400pt}}
\put(141,712){\makebox(0,0)[r]{90}}
\put(1419.0,712.0){\rule[-0.200pt]{4.818pt}{0.400pt}}
\put(161.0,777.0){\rule[-0.200pt]{4.818pt}{0.400pt}}
\put(141,777){\makebox(0,0)[r]{100}}
\put(1419.0,777.0){\rule[-0.200pt]{4.818pt}{0.400pt}}
\put(161.0,123.0){\rule[-0.200pt]{0.400pt}{4.818pt}}
\put(161,82){\makebox(0,0){85}}
\put(161.0,757.0){\rule[-0.200pt]{0.400pt}{4.818pt}}
\put(303.0,123.0){\rule[-0.200pt]{0.400pt}{4.818pt}}
\put(303,82){\makebox(0,0){90}}
\put(303.0,757.0){\rule[-0.200pt]{0.400pt}{4.818pt}}
\put(445.0,123.0){\rule[-0.200pt]{0.400pt}{4.818pt}}
\put(445,82){\makebox(0,0){95}}
\put(445.0,757.0){\rule[-0.200pt]{0.400pt}{4.818pt}}
\put(587.0,123.0){\rule[-0.200pt]{0.400pt}{4.818pt}}
\put(587,82){\makebox(0,0){100}}
\put(587.0,757.0){\rule[-0.200pt]{0.400pt}{4.818pt}}
\put(729.0,123.0){\rule[-0.200pt]{0.400pt}{4.818pt}}
\put(729,82){\makebox(0,0){105}}
\put(729.0,757.0){\rule[-0.200pt]{0.400pt}{4.818pt}}
\put(871.0,123.0){\rule[-0.200pt]{0.400pt}{4.818pt}}
\put(871,82){\makebox(0,0){110}}
\put(871.0,757.0){\rule[-0.200pt]{0.400pt}{4.818pt}}
\put(1013.0,123.0){\rule[-0.200pt]{0.400pt}{4.818pt}}
\put(1013,82){\makebox(0,0){115}}
\put(1013.0,757.0){\rule[-0.200pt]{0.400pt}{4.818pt}}
\put(1155.0,123.0){\rule[-0.200pt]{0.400pt}{4.818pt}}
\put(1155,82){\makebox(0,0){120}}
\put(1155.0,757.0){\rule[-0.200pt]{0.400pt}{4.818pt}}
\put(1297.0,123.0){\rule[-0.200pt]{0.400pt}{4.818pt}}
\put(1297,82){\makebox(0,0){125}}
\put(1297.0,757.0){\rule[-0.200pt]{0.400pt}{4.818pt}}
\put(1439.0,123.0){\rule[-0.200pt]{0.400pt}{4.818pt}}
\put(1439,82){\makebox(0,0){130}}
\put(1439.0,757.0){\rule[-0.200pt]{0.400pt}{4.818pt}}
\put(161.0,123.0){\rule[-0.200pt]{307.870pt}{0.400pt}}
\put(1439.0,123.0){\rule[-0.200pt]{0.400pt}{157.549pt}}
\put(161.0,777.0){\rule[-0.200pt]{307.870pt}{0.400pt}}
\put(40,450){\makebox(0,0){\shortstack{$\sigma$ \\ $[\textrm{fb}]$}}}
\put(800,21){\makebox(0,0){$m_{h^0} [\textrm{GeV/c}^2]$}}
\put(800,839){\makebox(0,0){$\sigma(\mu^- \mu^+ \rightarrow h^0 Z^0)$}}
\put(729,385){\makebox(0,0)[l]{$\sqrt{s} = 400 \textrm{GeV/c}, 
\tan\beta=50$}}
\put(161.0,123.0){\rule[-0.200pt]{0.400pt}{157.549pt}}
\put(1279,737){\makebox(0,0)[r]{ }}
\put(1299.0,737.0){\rule[-0.200pt]{24.090pt}{0.400pt}}
\put(274,125){\usebox{\plotpoint}}
\put(274,124.67){\rule{9.154pt}{0.400pt}}
\multiput(274.00,124.17)(19.000,1.000){2}{\rule{4.577pt}{0.400pt}}
\put(499,125.67){\rule{8.913pt}{0.400pt}}
\multiput(499.00,125.17)(18.500,1.000){2}{\rule{4.457pt}{0.400pt}}
\put(312.0,126.0){\rule[-0.200pt]{45.048pt}{0.400pt}}
\put(608,126.67){\rule{8.672pt}{0.400pt}}
\multiput(608.00,126.17)(18.000,1.000){2}{\rule{4.336pt}{0.400pt}}
\put(536.0,127.0){\rule[-0.200pt]{17.345pt}{0.400pt}}
\put(716,127.67){\rule{8.432pt}{0.400pt}}
\multiput(716.00,127.17)(17.500,1.000){2}{\rule{4.216pt}{0.400pt}}
\put(751,128.67){\rule{8.672pt}{0.400pt}}
\multiput(751.00,128.17)(18.000,1.000){2}{\rule{4.336pt}{0.400pt}}
\put(644.0,128.0){\rule[-0.200pt]{17.345pt}{0.400pt}}
\put(822,129.67){\rule{8.432pt}{0.400pt}}
\multiput(822.00,129.17)(17.500,1.000){2}{\rule{4.216pt}{0.400pt}}
\put(857,130.67){\rule{8.191pt}{0.400pt}}
\multiput(857.00,130.17)(17.000,1.000){2}{\rule{4.095pt}{0.400pt}}
\put(891,132.17){\rule{7.100pt}{0.400pt}}
\multiput(891.00,131.17)(20.264,2.000){2}{\rule{3.550pt}{0.400pt}}
\put(926,133.67){\rule{8.191pt}{0.400pt}}
\multiput(926.00,133.17)(17.000,1.000){2}{\rule{4.095pt}{0.400pt}}
\put(960,135.17){\rule{6.900pt}{0.400pt}}
\multiput(960.00,134.17)(19.679,2.000){2}{\rule{3.450pt}{0.400pt}}
\multiput(994.00,137.61)(7.383,0.447){3}{\rule{4.633pt}{0.108pt}}
\multiput(994.00,136.17)(24.383,3.000){2}{\rule{2.317pt}{0.400pt}}
\multiput(1028.00,140.61)(7.383,0.447){3}{\rule{4.633pt}{0.108pt}}
\multiput(1028.00,139.17)(24.383,3.000){2}{\rule{2.317pt}{0.400pt}}
\multiput(1062.00,143.60)(4.722,0.468){5}{\rule{3.400pt}{0.113pt}}
\multiput(1062.00,142.17)(25.943,4.000){2}{\rule{1.700pt}{0.400pt}}
\multiput(1095.00,147.59)(2.932,0.482){9}{\rule{2.300pt}{0.116pt}}
\multiput(1095.00,146.17)(28.226,6.000){2}{\rule{1.150pt}{0.400pt}}
\multiput(1128.00,153.59)(2.145,0.488){13}{\rule{1.750pt}{0.117pt}}
\multiput(1128.00,152.17)(29.368,8.000){2}{\rule{0.875pt}{0.400pt}}
\multiput(1161.00,161.58)(1.210,0.493){23}{\rule{1.054pt}{0.119pt}}
\multiput(1161.00,160.17)(28.813,13.000){2}{\rule{0.527pt}{0.400pt}}
\multiput(1192.00,174.58)(0.866,0.495){33}{\rule{0.789pt}{0.119pt}}
\multiput(1192.00,173.17)(29.363,18.000){2}{\rule{0.394pt}{0.400pt}}
\multiput(1223.00,192.58)(0.535,0.497){53}{\rule{0.529pt}{0.120pt}}
\multiput(1223.00,191.17)(28.903,28.000){2}{\rule{0.264pt}{0.400pt}}
\multiput(1253.58,220.00)(0.497,0.806){53}{\rule{0.120pt}{0.743pt}}
\multiput(1252.17,220.00)(28.000,43.458){2}{\rule{0.400pt}{0.371pt}}
\multiput(1281.58,265.00)(0.496,1.408){45}{\rule{0.120pt}{1.217pt}}
\multiput(1280.17,265.00)(24.000,64.475){2}{\rule{0.400pt}{0.608pt}}
\multiput(1305.58,332.00)(0.496,2.204){37}{\rule{0.119pt}{1.840pt}}
\multiput(1304.17,332.00)(20.000,83.181){2}{\rule{0.400pt}{0.920pt}}
\multiput(1325.58,419.00)(0.494,3.097){27}{\rule{0.119pt}{2.527pt}}
\multiput(1324.17,419.00)(15.000,85.756){2}{\rule{0.400pt}{1.263pt}}
\multiput(1340.58,510.00)(0.492,3.420){19}{\rule{0.118pt}{2.755pt}}
\multiput(1339.17,510.00)(11.000,67.283){2}{\rule{0.400pt}{1.377pt}}
\multiput(1351.59,583.00)(0.485,3.696){11}{\rule{0.117pt}{2.900pt}}
\multiput(1350.17,583.00)(7.000,42.981){2}{\rule{0.400pt}{1.450pt}}
\multiput(1358.59,632.00)(0.477,3.493){7}{\rule{0.115pt}{2.660pt}}
\multiput(1357.17,632.00)(5.000,26.479){2}{\rule{0.400pt}{1.330pt}}
\multiput(1363.61,664.00)(0.447,4.258){3}{\rule{0.108pt}{2.767pt}}
\multiput(1362.17,664.00)(3.000,14.258){2}{\rule{0.400pt}{1.383pt}}
\put(1366.17,684){\rule{0.400pt}{2.700pt}}
\multiput(1365.17,684.00)(2.000,7.396){2}{\rule{0.400pt}{1.350pt}}
\put(1368.17,697){\rule{0.400pt}{1.700pt}}
\multiput(1367.17,697.00)(2.000,4.472){2}{\rule{0.400pt}{0.850pt}}
\put(1370.17,705){\rule{0.400pt}{1.300pt}}
\multiput(1369.17,705.00)(2.000,3.302){2}{\rule{0.400pt}{0.650pt}}
\put(1371.67,711){\rule{0.400pt}{1.204pt}}
\multiput(1371.17,711.00)(1.000,2.500){2}{\rule{0.400pt}{0.602pt}}
\put(1372.67,716){\rule{0.400pt}{0.723pt}}
\multiput(1372.17,716.00)(1.000,1.500){2}{\rule{0.400pt}{0.361pt}}
\put(1373.67,719){\rule{0.400pt}{0.482pt}}
\multiput(1373.17,719.00)(1.000,1.000){2}{\rule{0.400pt}{0.241pt}}
\put(787.0,130.0){\rule[-0.200pt]{8.431pt}{0.400pt}}
\put(1374.67,723){\rule{0.400pt}{0.482pt}}
\multiput(1374.17,723.00)(1.000,1.000){2}{\rule{0.400pt}{0.241pt}}
\put(1375.0,721.0){\rule[-0.200pt]{0.400pt}{0.482pt}}
\put(1376,725.67){\rule{0.241pt}{0.400pt}}
\multiput(1376.00,725.17)(0.500,1.000){2}{\rule{0.120pt}{0.400pt}}
\put(1376.0,725.0){\usebox{\plotpoint}}
\put(1377,727){\usebox{\plotpoint}}
\put(1377,727.67){\rule{0.241pt}{0.400pt}}
\multiput(1377.00,727.17)(0.500,1.000){2}{\rule{0.120pt}{0.400pt}}
\put(1377.0,727.0){\usebox{\plotpoint}}
\put(1378,729){\usebox{\plotpoint}}
\put(1378,729){\usebox{\plotpoint}}
\put(1378.0,729.0){\usebox{\plotpoint}}
\put(1378.0,730.0){\usebox{\plotpoint}}
\put(1379.0,730.0){\usebox{\plotpoint}}
\put(1379.0,731.0){\usebox{\plotpoint}}
\put(1380.0,731.0){\usebox{\plotpoint}}
\put(1380.0,732.0){\rule[-0.200pt]{0.482pt}{0.400pt}}
\put(1382.0,732.0){\usebox{\plotpoint}}
\end{picture}
\caption{Total cross section for the process
$\mu^- \mu^+ \rightarrow h^0 Z^0$ as a function
of $m_{h^0}$. We have taken
$\sqrt{s} = 400 \textrm{GeV/c}$ and $\tan\beta= 50$.}
\label{mumuhz_graph2}
\end{center}
\end{figure}
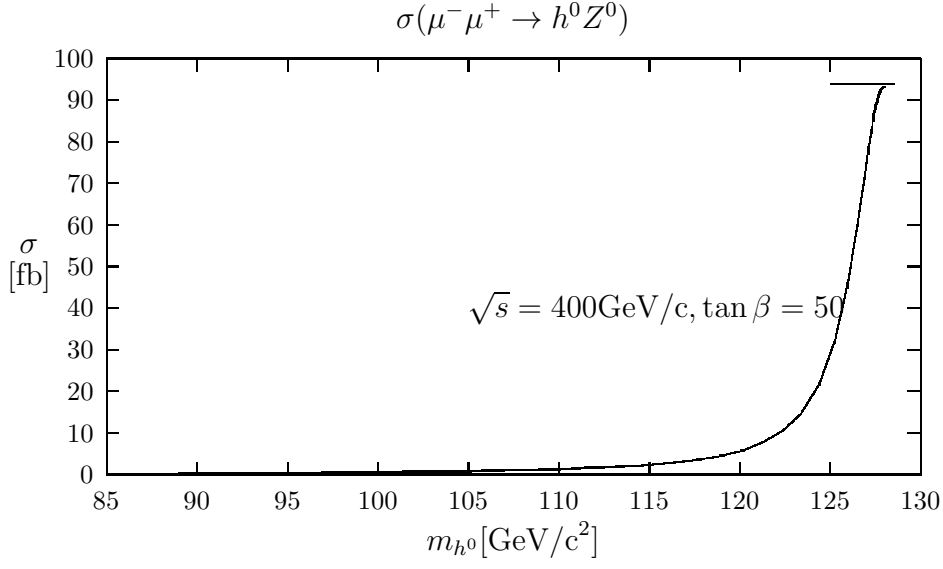

In Figures \ref{mumuhz_graph1} and \ref{mumuhz_graph2}
, the total cross section for 
$\mu^- \mu^+ \rightarrow h^0 Z^0$, is plotted as a function
$m_{h^0}$ for several values of  
$\sqrt{s}$ and $\tan\beta$. These total cross sections were
plotted considering the radiative corrections of the masses
given by Equations
(\ref{mhoGrc}), (\ref{Grc}), (\ref{tan_alpharc}) and (\ref{Frc}). 
According to these graphs, the total cross section becomes important
in the mass interval $118 \leq m_{h^0} \leq 128 [\textrm{GeV/c}^2]$.

The Standard Model cross section is:

\begin{eqnarray}
\sigma (\mu^- \mu^+ \rightarrow h_{SM}^0 Z^0)_{SM} = \frac{G_F^2 m_Z^4
}
{48 \pi s^2}
\frac{\left( 1 - 4 \sin^2 \theta_W + 8 \sin^4 \theta_W \right)}
{\left[ \left( s - m_Z^2 \right)^2 + m_Z^2 \Gamma_Z^2 \right]}
\nonumber \\
\times \left[ 12 s m_Z^2 + 
\Lambda ( s, m_{h_{SM}^0}^2, m_Z^2 ) \right]
\nonumber \\ 
\times\Lambda^{1/2} ( s, m_{h_{SM}^0}^2, m_Z^2 ) \times
\left( 3.8938 \times 10^{11} \right) \textrm{fb}
\label{sigmaSM_muhz}
\end{eqnarray}

\noindent where $h_{SM}$ is the Standard Model higgs boson.
 
\begin{figure}
\begin{center}
\input{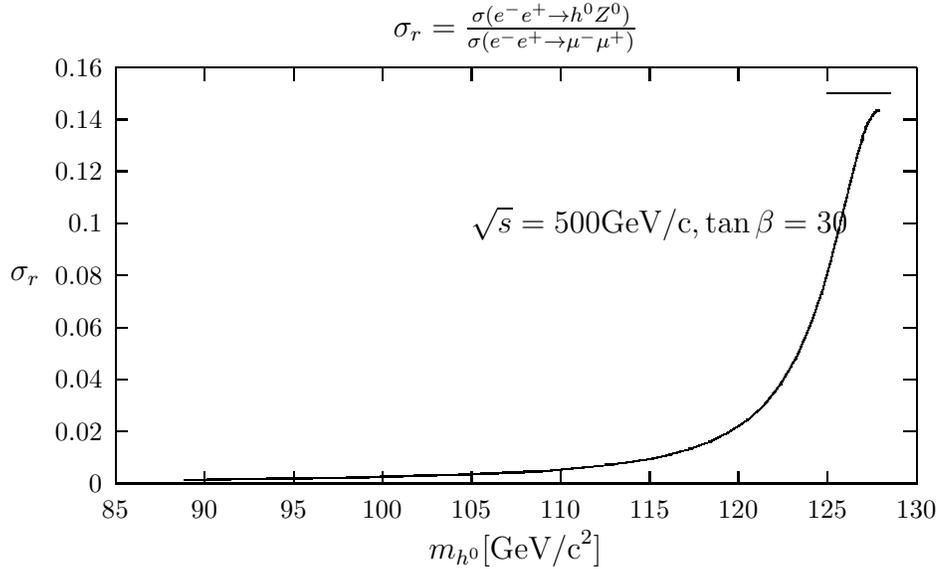}
\caption{Total cross section
$\sigma(e^- e^+ \rightarrow h^0 Z^0)$ compared with the
cross section $\sigma(e^- e^+ \rightarrow
\mu^- \mu^+)$ as a function
of $m_{h^0}$. We have taken
$\sqrt{s} = 500 \textrm{GeV/c}$ and $\tan\beta= 30$.}
\label{annihi1}
\end{center}
\end{figure}

The production  cross section corresponding to
$e^- e^+ \rightarrow h^0 Z^0$
is given by an expression identical to (\ref{sigma_muhz}).
In terms of the cross section
$\sigma \left( e^- e^+ \rightarrow \mu^- \mu^+ \right)$ we can write:
\begin{eqnarray}
\lefteqn{
\frac{\sigma \left( e^- e^+ \rightarrow h^0 Z^0 \right) }
{\sigma \left( e^- e^+ \rightarrow \mu^- \mu^+ \right)} =
\frac{1}{128 s} \frac{\left( \tan\beta - \tan\alpha \right)
^2}{\left( 1 + \tan^2\alpha \right) \left( 1 + 
\tan^2\beta \right) } \Lambda^{1/2} ( s, m_{h^0}^2, m_Z^{2} ) 
}
\nonumber \\ & &
\times
\frac{ \left( 1 - 4 \sin^2\theta_W
+ 8 \sin^4\theta_W \right)}{\sin^4\theta_W 
\left( 1 - \sin^2\theta_W \right)^2}
 \frac{\left[ 12 s m_Z^2 + \Lambda ( s, m_{h^0}^2, m_Z^2 )
\right] }
{\left[ \left( s - m_Z^2 \right)^2 + m_Z^2 \Gamma_Z^2
\right]}
\label{annihilation1}
\end{eqnarray} 
Equation (\ref{annihilation1}) is plotted in Figure 
\ref{annihi1}, as a function of
 $m_{h^0}$ for $\sqrt{s} = 500 \textrm{GeV/c}$ and $\tan\beta= 30$.

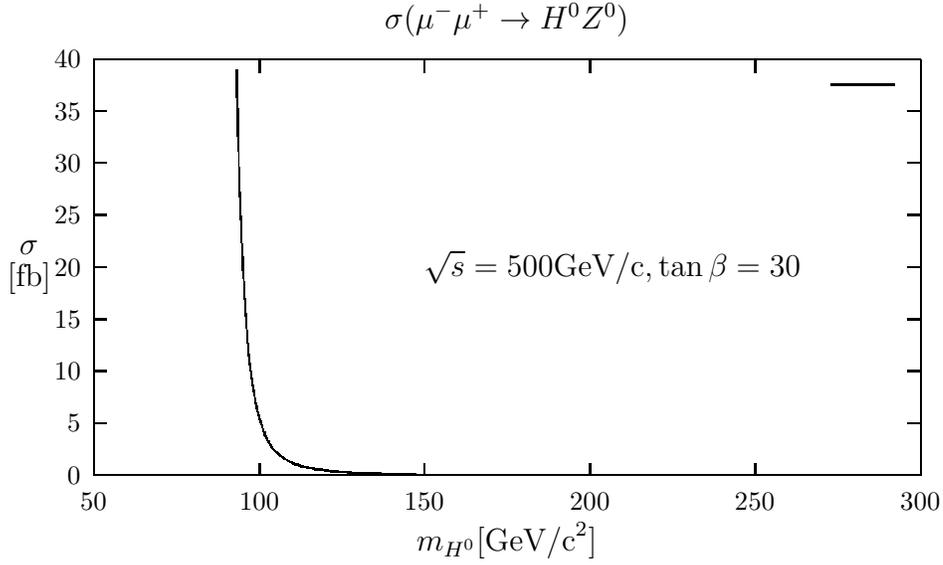
\begin{figure}
\begin{center}
% GNUPLOT: LaTeX picture
\setlength{\unitlength}{0.240900pt}
\ifx\plotpoint\undefined\newsavebox{\plotpoint}\fi
\sbox{\plotpoint}{\rule[-0.200pt]{0.400pt}{0.400pt}}%
\begin{picture}(1500,900)(0,0)
\font\gnuplot=cmr10 at 10pt
\gnuplot
\sbox{\plotpoint}{\rule[-0.200pt]{0.400pt}{0.400pt}}%
\put(141.0,123.0){\rule[-0.200pt]{4.818pt}{0.400pt}}
\put(121,123){\makebox(0,0)[r]{0}}
\put(1419.0,123.0){\rule[-0.200pt]{4.818pt}{0.400pt}}
\put(141.0,205.0){\rule[-0.200pt]{4.818pt}{0.400pt}}
\put(121,205){\makebox(0,0)[r]{5}}
\put(1419.0,205.0){\rule[-0.200pt]{4.818pt}{0.400pt}}
\put(141.0,287.0){\rule[-0.200pt]{4.818pt}{0.400pt}}
\put(121,287){\makebox(0,0)[r]{10}}
\put(1419.0,287.0){\rule[-0.200pt]{4.818pt}{0.400pt}}
\put(141.0,368.0){\rule[-0.200pt]{4.818pt}{0.400pt}}
\put(121,368){\makebox(0,0)[r]{15}}
\put(1419.0,368.0){\rule[-0.200pt]{4.818pt}{0.400pt}}
\put(141.0,450.0){\rule[-0.200pt]{4.818pt}{0.400pt}}
\put(121,450){\makebox(0,0)[r]{20}}
\put(1419.0,450.0){\rule[-0.200pt]{4.818pt}{0.400pt}}
\put(141.0,532.0){\rule[-0.200pt]{4.818pt}{0.400pt}}
\put(121,532){\makebox(0,0)[r]{25}}
\put(1419.0,532.0){\rule[-0.200pt]{4.818pt}{0.400pt}}
\put(141.0,614.0){\rule[-0.200pt]{4.818pt}{0.400pt}}
\put(121,614){\makebox(0,0)[r]{30}}
\put(1419.0,614.0){\rule[-0.200pt]{4.818pt}{0.400pt}}
\put(141.0,695.0){\rule[-0.200pt]{4.818pt}{0.400pt}}
\put(121,695){\makebox(0,0)[r]{35}}
\put(1419.0,695.0){\rule[-0.200pt]{4.818pt}{0.400pt}}
\put(141.0,777.0){\rule[-0.200pt]{4.818pt}{0.400pt}}
\put(121,777){\makebox(0,0)[r]{40}}
\put(1419.0,777.0){\rule[-0.200pt]{4.818pt}{0.400pt}}
\put(141.0,123.0){\rule[-0.200pt]{0.400pt}{4.818pt}}
\put(141,82){\makebox(0,0){50}}
\put(141.0,757.0){\rule[-0.200pt]{0.400pt}{4.818pt}}
\put(401.0,123.0){\rule[-0.200pt]{0.400pt}{4.818pt}}
\put(401,82){\makebox(0,0){100}}
\put(401.0,757.0){\rule[-0.200pt]{0.400pt}{4.818pt}}
\put(660.0,123.0){\rule[-0.200pt]{0.400pt}{4.818pt}}
\put(660,82){\makebox(0,0){150}}
\put(660.0,757.0){\rule[-0.200pt]{0.400pt}{4.818pt}}
\put(920.0,123.0){\rule[-0.200pt]{0.400pt}{4.818pt}}
\put(920,82){\makebox(0,0){200}}
\put(920.0,757.0){\rule[-0.200pt]{0.400pt}{4.818pt}}
\put(1179.0,123.0){\rule[-0.200pt]{0.400pt}{4.818pt}}
\put(1179,82){\makebox(0,0){250}}
\put(1179.0,757.0){\rule[-0.200pt]{0.400pt}{4.818pt}}
\put(1439.0,123.0){\rule[-0.200pt]{0.400pt}{4.818pt}}
\put(1439,82){\makebox(0,0){300}}
\put(1439.0,757.0){\rule[-0.200pt]{0.400pt}{4.818pt}}
\put(141.0,123.0){\rule[-0.200pt]{312.688pt}{0.400pt}}
\put(1439.0,123.0){\rule[-0.200pt]{0.400pt}{157.549pt}}
\put(141.0,777.0){\rule[-0.200pt]{312.688pt}{0.400pt}}
\put(40,450){\makebox(0,0){\shortstack{$\sigma$ \\ $[\textrm{fb}]$}}}
\put(790,21){\makebox(0,0){$m_{H^0} [\textrm{GeV/c}^2]$}}
\put(790,839){\makebox(0,0){$\sigma(\mu^- \mu^+ \rightarrow H^0 Z^0)$}}
\put(660,450){\makebox(0,0)[l]{$\sqrt{s} = 500 \textrm{GeV/c}, 
\tan\beta=30$}}
\put(141.0,123.0){\rule[-0.200pt]{0.400pt}{157.549pt}}
\put(1279,737){\makebox(0,0)[r]{ }}
\put(1299.0,737.0){\rule[-0.200pt]{24.090pt}{0.400pt}}
\put(366,760){\usebox{\plotpoint}}
\put(366.17,655){\rule{0.400pt}{21.100pt}}
\multiput(365.17,716.21)(2.000,-61.206){2}{\rule{0.400pt}{10.550pt}}
\multiput(368.60,608.92)(0.468,-15.981){5}{\rule{0.113pt}{11.100pt}}
\multiput(367.17,631.96)(4.000,-86.961){2}{\rule{0.400pt}{5.550pt}}
\multiput(372.60,503.07)(0.468,-14.518){5}{\rule{0.113pt}{10.100pt}}
\multiput(371.17,524.04)(4.000,-79.037){2}{\rule{0.400pt}{5.050pt}}
\multiput(376.59,417.69)(0.477,-8.948){7}{\rule{0.115pt}{6.580pt}}
\multiput(375.17,431.34)(5.000,-67.343){2}{\rule{0.400pt}{3.290pt}}
\multiput(381.59,343.33)(0.477,-6.721){7}{\rule{0.115pt}{4.980pt}}
\multiput(380.17,353.66)(5.000,-50.664){2}{\rule{0.400pt}{2.490pt}}
\multiput(386.59,290.41)(0.482,-3.926){9}{\rule{0.116pt}{3.033pt}}
\multiput(385.17,296.70)(6.000,-37.704){2}{\rule{0.400pt}{1.517pt}}
\multiput(392.59,249.73)(0.482,-2.841){9}{\rule{0.116pt}{2.233pt}}
\multiput(391.17,254.36)(6.000,-27.365){2}{\rule{0.400pt}{1.117pt}}
\multiput(398.59,220.50)(0.482,-1.937){9}{\rule{0.116pt}{1.567pt}}
\multiput(397.17,223.75)(6.000,-18.748){2}{\rule{0.400pt}{0.783pt}}
\multiput(404.59,199.88)(0.482,-1.485){9}{\rule{0.116pt}{1.233pt}}
\multiput(403.17,202.44)(6.000,-14.440){2}{\rule{0.400pt}{0.617pt}}
\multiput(410.59,184.26)(0.482,-1.033){9}{\rule{0.116pt}{0.900pt}}
\multiput(409.17,186.13)(6.000,-10.132){2}{\rule{0.400pt}{0.450pt}}
\multiput(416.59,172.82)(0.482,-0.852){9}{\rule{0.116pt}{0.767pt}}
\multiput(415.17,174.41)(6.000,-8.409){2}{\rule{0.400pt}{0.383pt}}
\multiput(422.59,163.65)(0.482,-0.581){9}{\rule{0.116pt}{0.567pt}}
\multiput(421.17,164.82)(6.000,-5.824){2}{\rule{0.400pt}{0.283pt}}
\multiput(428.00,157.93)(0.581,-0.482){9}{\rule{0.567pt}{0.116pt}}
\multiput(428.00,158.17)(5.824,-6.000){2}{\rule{0.283pt}{0.400pt}}
\multiput(435.00,151.94)(0.774,-0.468){5}{\rule{0.700pt}{0.113pt}}
\multiput(435.00,152.17)(4.547,-4.000){2}{\rule{0.350pt}{0.400pt}}
\multiput(441.00,147.94)(0.774,-0.468){5}{\rule{0.700pt}{0.113pt}}
\multiput(441.00,148.17)(4.547,-4.000){2}{\rule{0.350pt}{0.400pt}}
\multiput(447.00,143.95)(1.355,-0.447){3}{\rule{1.033pt}{0.108pt}}
\multiput(447.00,144.17)(4.855,-3.000){2}{\rule{0.517pt}{0.400pt}}
\put(454,140.17){\rule{1.300pt}{0.400pt}}
\multiput(454.00,141.17)(3.302,-2.000){2}{\rule{0.650pt}{0.400pt}}
\put(460,138.17){\rule{1.300pt}{0.400pt}}
\multiput(460.00,139.17)(3.302,-2.000){2}{\rule{0.650pt}{0.400pt}}
\put(466,136.17){\rule{1.300pt}{0.400pt}}
\multiput(466.00,137.17)(3.302,-2.000){2}{\rule{0.650pt}{0.400pt}}
\put(472,134.67){\rule{1.686pt}{0.400pt}}
\multiput(472.00,135.17)(3.500,-1.000){2}{\rule{0.843pt}{0.400pt}}
\put(479,133.67){\rule{1.445pt}{0.400pt}}
\multiput(479.00,134.17)(3.000,-1.000){2}{\rule{0.723pt}{0.400pt}}
\put(485,132.17){\rule{1.300pt}{0.400pt}}
\multiput(485.00,133.17)(3.302,-2.000){2}{\rule{0.650pt}{0.400pt}}
\put(497,130.67){\rule{1.445pt}{0.400pt}}
\multiput(497.00,131.17)(3.000,-1.000){2}{\rule{0.723pt}{0.400pt}}
\put(503,129.67){\rule{1.686pt}{0.400pt}}
\multiput(503.00,130.17)(3.500,-1.000){2}{\rule{0.843pt}{0.400pt}}
\put(510,128.67){\rule{1.445pt}{0.400pt}}
\multiput(510.00,129.17)(3.000,-1.000){2}{\rule{0.723pt}{0.400pt}}
\put(491.0,132.0){\rule[-0.200pt]{1.445pt}{0.400pt}}
\put(522,127.67){\rule{1.445pt}{0.400pt}}
\multiput(522.00,128.17)(3.000,-1.000){2}{\rule{0.723pt}{0.400pt}}
\put(516.0,129.0){\rule[-0.200pt]{1.445pt}{0.400pt}}
\put(540,126.67){\rule{1.445pt}{0.400pt}}
\multiput(540.00,127.17)(3.000,-1.000){2}{\rule{0.723pt}{0.400pt}}
\put(528.0,128.0){\rule[-0.200pt]{2.891pt}{0.400pt}}
\put(559,125.67){\rule{1.445pt}{0.400pt}}
\multiput(559.00,126.17)(3.000,-1.000){2}{\rule{0.723pt}{0.400pt}}
\put(546.0,127.0){\rule[-0.200pt]{3.132pt}{0.400pt}}
\put(589,124.67){\rule{1.445pt}{0.400pt}}
\multiput(589.00,125.17)(3.000,-1.000){2}{\rule{0.723pt}{0.400pt}}
\put(565.0,126.0){\rule[-0.200pt]{5.782pt}{0.400pt}}
\put(642,123.67){\rule{1.445pt}{0.400pt}}
\multiput(642.00,124.17)(3.000,-1.000){2}{\rule{0.723pt}{0.400pt}}
\put(595.0,125.0){\rule[-0.200pt]{11.322pt}{0.400pt}}
\put(782,122.67){\rule{1.204pt}{0.400pt}}
\multiput(782.00,123.17)(2.500,-1.000){2}{\rule{0.602pt}{0.400pt}}
\put(648.0,124.0){\rule[-0.200pt]{32.281pt}{0.400pt}}
\put(787.0,123.0){\rule[-0.200pt]{153.694pt}{0.400pt}}
\end{picture}
\caption{Total cross section for the process
$\mu^- \mu^+ \rightarrow H^0 Z^0$ as a function
of $m_{H^0}$. The radiative corrections of the masses were not 
taken into account.
We have taken
$\sqrt{s} = 500 \textrm{GeV/c}$ and $\tan\beta= 30$.}
\label{mumu_HZ1}
\end{center}
\end{figure}

\begin{figure}
\begin{center}
% GNUPLOT: LaTeX picture
\setlength{\unitlength}{0.240900pt}
\ifx\plotpoint\undefined\newsavebox{\plotpoint}\fi
\sbox{\plotpoint}{\rule[-0.200pt]{0.400pt}{0.400pt}}%
\begin{picture}(1500,900)(0,0)
\font\gnuplot=cmr10 at 10pt
\gnuplot
\sbox{\plotpoint}{\rule[-0.200pt]{0.400pt}{0.400pt}}%
\put(141.0,123.0){\rule[-0.200pt]{4.818pt}{0.400pt}}
\put(121,123){\makebox(0,0)[r]{0}}
\put(1419.0,123.0){\rule[-0.200pt]{4.818pt}{0.400pt}}
\put(141.0,232.0){\rule[-0.200pt]{4.818pt}{0.400pt}}
\put(121,232){\makebox(0,0)[r]{10}}
\put(1419.0,232.0){\rule[-0.200pt]{4.818pt}{0.400pt}}
\put(141.0,341.0){\rule[-0.200pt]{4.818pt}{0.400pt}}
\put(121,341){\makebox(0,0)[r]{20}}
\put(1419.0,341.0){\rule[-0.200pt]{4.818pt}{0.400pt}}
\put(141.0,450.0){\rule[-0.200pt]{4.818pt}{0.400pt}}
\put(121,450){\makebox(0,0)[r]{30}}
\put(1419.0,450.0){\rule[-0.200pt]{4.818pt}{0.400pt}}
\put(141.0,559.0){\rule[-0.200pt]{4.818pt}{0.400pt}}
\put(121,559){\makebox(0,0)[r]{40}}
\put(1419.0,559.0){\rule[-0.200pt]{4.818pt}{0.400pt}}
\put(141.0,668.0){\rule[-0.200pt]{4.818pt}{0.400pt}}
\put(121,668){\makebox(0,0)[r]{50}}
\put(1419.0,668.0){\rule[-0.200pt]{4.818pt}{0.400pt}}
\put(141.0,777.0){\rule[-0.200pt]{4.818pt}{0.400pt}}
\put(121,777){\makebox(0,0)[r]{60}}
\put(1419.0,777.0){\rule[-0.200pt]{4.818pt}{0.400pt}}
\put(141.0,123.0){\rule[-0.200pt]{0.400pt}{4.818pt}}
\put(141,82){\makebox(0,0){120}}
\put(141.0,757.0){\rule[-0.200pt]{0.400pt}{4.818pt}}
\put(326.0,123.0){\rule[-0.200pt]{0.400pt}{4.818pt}}
\put(326,82){\makebox(0,0){140}}
\put(326.0,757.0){\rule[-0.200pt]{0.400pt}{4.818pt}}
\put(512.0,123.0){\rule[-0.200pt]{0.400pt}{4.818pt}}
\put(512,82){\makebox(0,0){160}}
\put(512.0,757.0){\rule[-0.200pt]{0.400pt}{4.818pt}}
\put(697.0,123.0){\rule[-0.200pt]{0.400pt}{4.818pt}}
\put(697,82){\makebox(0,0){180}}
\put(697.0,757.0){\rule[-0.200pt]{0.400pt}{4.818pt}}
\put(883.0,123.0){\rule[-0.200pt]{0.400pt}{4.818pt}}
\put(883,82){\makebox(0,0){200}}
\put(883.0,757.0){\rule[-0.200pt]{0.400pt}{4.818pt}}
\put(1068.0,123.0){\rule[-0.200pt]{0.400pt}{4.818pt}}
\put(1068,82){\makebox(0,0){220}}
\put(1068.0,757.0){\rule[-0.200pt]{0.400pt}{4.818pt}}
\put(1254.0,123.0){\rule[-0.200pt]{0.400pt}{4.818pt}}
\put(1254,82){\makebox(0,0){240}}
\put(1254.0,757.0){\rule[-0.200pt]{0.400pt}{4.818pt}}
\put(1439.0,123.0){\rule[-0.200pt]{0.400pt}{4.818pt}}
\put(1439,82){\makebox(0,0){260}}
\put(1439.0,757.0){\rule[-0.200pt]{0.400pt}{4.818pt}}
\put(141.0,123.0){\rule[-0.200pt]{312.688pt}{0.400pt}}
\put(1439.0,123.0){\rule[-0.200pt]{0.400pt}{157.549pt}}
\put(141.0,777.0){\rule[-0.200pt]{312.688pt}{0.400pt}}
\put(40,450){\makebox(0,0){\shortstack{$\sigma$ \\ $[\textrm{fb}]$}}}
\put(790,21){\makebox(0,0){$m_{H^0} [\textrm{GeV/c}^2]$}}
\put(790,839){\makebox(0,0){$\sigma(\mu^- \mu^+ \rightarrow H^0 Z^0)$}}
\put(419,341){\makebox(0,0)[l]{$\sqrt{s} = 500 \textrm{GeV/c}, \tan\beta = 
30$}}
\put(141.0,123.0){\rule[-0.200pt]{0.400pt}{157.549pt}}
\put(1279,737){\makebox(0,0)[r]{ }}
\put(1299.0,737.0){\rule[-0.200pt]{24.090pt}{0.400pt}}
\put(217,731){\usebox{\plotpoint}}
\put(217,731){\usebox{\plotpoint}}
\put(217.0,730.0){\usebox{\plotpoint}}
\put(217.0,730.0){\usebox{\plotpoint}}
\put(218,723.67){\rule{0.241pt}{0.400pt}}
\multiput(218.00,724.17)(0.500,-1.000){2}{\rule{0.120pt}{0.400pt}}
\put(218.0,725.0){\rule[-0.200pt]{0.400pt}{1.204pt}}
\put(218.67,718){\rule{0.400pt}{0.482pt}}
\multiput(218.17,719.00)(1.000,-1.000){2}{\rule{0.400pt}{0.241pt}}
\put(219.0,720.0){\rule[-0.200pt]{0.400pt}{0.964pt}}
\put(219.67,710){\rule{0.400pt}{0.723pt}}
\multiput(219.17,711.50)(1.000,-1.500){2}{\rule{0.400pt}{0.361pt}}
\put(220.0,713.0){\rule[-0.200pt]{0.400pt}{1.204pt}}
\put(220.67,702){\rule{0.400pt}{0.964pt}}
\multiput(220.17,704.00)(1.000,-2.000){2}{\rule{0.400pt}{0.482pt}}
\put(221.0,706.0){\rule[-0.200pt]{0.400pt}{0.964pt}}
\put(221.67,689){\rule{0.400pt}{1.686pt}}
\multiput(221.17,692.50)(1.000,-3.500){2}{\rule{0.400pt}{0.843pt}}
\put(222.67,680){\rule{0.400pt}{2.168pt}}
\multiput(222.17,684.50)(1.000,-4.500){2}{\rule{0.400pt}{1.084pt}}
\put(223.67,668){\rule{0.400pt}{2.891pt}}
\multiput(223.17,674.00)(1.000,-6.000){2}{\rule{0.400pt}{1.445pt}}
\put(224.67,653){\rule{0.400pt}{3.614pt}}
\multiput(224.17,660.50)(1.000,-7.500){2}{\rule{0.400pt}{1.807pt}}
\put(226.17,633){\rule{0.400pt}{4.100pt}}
\multiput(225.17,644.49)(2.000,-11.490){2}{\rule{0.400pt}{2.050pt}}
\put(228.17,607){\rule{0.400pt}{5.300pt}}
\multiput(227.17,622.00)(2.000,-15.000){2}{\rule{0.400pt}{2.650pt}}
\put(230.17,573){\rule{0.400pt}{6.900pt}}
\multiput(229.17,592.68)(2.000,-19.679){2}{\rule{0.400pt}{3.450pt}}
\multiput(232.61,548.79)(0.447,-9.393){3}{\rule{0.108pt}{5.833pt}}
\multiput(231.17,560.89)(3.000,-30.893){2}{\rule{0.400pt}{2.917pt}}
\multiput(235.60,508.83)(0.468,-7.207){5}{\rule{0.113pt}{5.100pt}}
\multiput(234.17,519.41)(4.000,-39.415){2}{\rule{0.400pt}{2.550pt}}
\multiput(239.59,460.99)(0.477,-6.165){7}{\rule{0.115pt}{4.580pt}}
\multiput(238.17,470.49)(5.000,-46.494){2}{\rule{0.400pt}{2.290pt}}
\multiput(244.59,408.36)(0.482,-4.921){9}{\rule{0.116pt}{3.767pt}}
\multiput(243.17,416.18)(6.000,-47.182){2}{\rule{0.400pt}{1.883pt}}
\multiput(250.59,356.72)(0.485,-3.772){11}{\rule{0.117pt}{2.957pt}}
\multiput(249.17,362.86)(7.000,-43.862){2}{\rule{0.400pt}{1.479pt}}
\multiput(257.59,308.39)(0.485,-3.239){11}{\rule{0.117pt}{2.557pt}}
\multiput(256.17,313.69)(7.000,-37.693){2}{\rule{0.400pt}{1.279pt}}
\multiput(264.59,269.50)(0.489,-1.893){15}{\rule{0.118pt}{1.567pt}}
\multiput(263.17,272.75)(9.000,-29.748){2}{\rule{0.400pt}{0.783pt}}
\multiput(273.59,237.79)(0.489,-1.485){15}{\rule{0.118pt}{1.256pt}}
\multiput(272.17,240.39)(9.000,-23.394){2}{\rule{0.400pt}{0.628pt}}
\multiput(282.59,213.08)(0.489,-1.077){15}{\rule{0.118pt}{0.944pt}}
\multiput(281.17,215.04)(9.000,-17.040){2}{\rule{0.400pt}{0.472pt}}
\multiput(291.59,194.82)(0.489,-0.844){15}{\rule{0.118pt}{0.767pt}}
\multiput(290.17,196.41)(9.000,-13.409){2}{\rule{0.400pt}{0.383pt}}
\multiput(300.58,180.76)(0.491,-0.547){17}{\rule{0.118pt}{0.540pt}}
\multiput(299.17,181.88)(10.000,-9.879){2}{\rule{0.400pt}{0.270pt}}
\multiput(310.00,170.93)(0.626,-0.488){13}{\rule{0.600pt}{0.117pt}}
\multiput(310.00,171.17)(8.755,-8.000){2}{\rule{0.300pt}{0.400pt}}
\multiput(320.00,162.93)(0.721,-0.485){11}{\rule{0.671pt}{0.117pt}}
\multiput(320.00,163.17)(8.606,-7.000){2}{\rule{0.336pt}{0.400pt}}
\multiput(330.00,155.93)(1.044,-0.477){7}{\rule{0.900pt}{0.115pt}}
\multiput(330.00,156.17)(8.132,-5.000){2}{\rule{0.450pt}{0.400pt}}
\multiput(340.00,150.93)(1.044,-0.477){7}{\rule{0.900pt}{0.115pt}}
\multiput(340.00,151.17)(8.132,-5.000){2}{\rule{0.450pt}{0.400pt}}
\multiput(350.00,145.95)(2.025,-0.447){3}{\rule{1.433pt}{0.108pt}}
\multiput(350.00,146.17)(7.025,-3.000){2}{\rule{0.717pt}{0.400pt}}
\multiput(360.00,142.95)(2.025,-0.447){3}{\rule{1.433pt}{0.108pt}}
\multiput(360.00,143.17)(7.025,-3.000){2}{\rule{0.717pt}{0.400pt}}
\put(370,139.17){\rule{2.300pt}{0.400pt}}
\multiput(370.00,140.17)(6.226,-2.000){2}{\rule{1.150pt}{0.400pt}}
\put(381,137.17){\rule{2.100pt}{0.400pt}}
\multiput(381.00,138.17)(5.641,-2.000){2}{\rule{1.050pt}{0.400pt}}
\put(391,135.17){\rule{2.100pt}{0.400pt}}
\multiput(391.00,136.17)(5.641,-2.000){2}{\rule{1.050pt}{0.400pt}}
\put(401,133.67){\rule{2.650pt}{0.400pt}}
\multiput(401.00,134.17)(5.500,-1.000){2}{\rule{1.325pt}{0.400pt}}
\put(412,132.67){\rule{2.409pt}{0.400pt}}
\multiput(412.00,133.17)(5.000,-1.000){2}{\rule{1.204pt}{0.400pt}}
\put(422,131.67){\rule{2.409pt}{0.400pt}}
\multiput(422.00,132.17)(5.000,-1.000){2}{\rule{1.204pt}{0.400pt}}
\put(432,130.67){\rule{2.409pt}{0.400pt}}
\multiput(432.00,131.17)(5.000,-1.000){2}{\rule{1.204pt}{0.400pt}}
\put(442,129.67){\rule{2.650pt}{0.400pt}}
\multiput(442.00,130.17)(5.500,-1.000){2}{\rule{1.325pt}{0.400pt}}
\put(222.0,696.0){\rule[-0.200pt]{0.400pt}{1.445pt}}
\put(463,128.67){\rule{2.409pt}{0.400pt}}
\multiput(463.00,129.17)(5.000,-1.000){2}{\rule{1.204pt}{0.400pt}}
\put(473,127.67){\rule{2.650pt}{0.400pt}}
\multiput(473.00,128.17)(5.500,-1.000){2}{\rule{1.325pt}{0.400pt}}
\put(453.0,130.0){\rule[-0.200pt]{2.409pt}{0.400pt}}
\put(504,126.67){\rule{2.650pt}{0.400pt}}
\multiput(504.00,127.17)(5.500,-1.000){2}{\rule{1.325pt}{0.400pt}}
\put(484.0,128.0){\rule[-0.200pt]{4.818pt}{0.400pt}}
\put(535,125.67){\rule{2.409pt}{0.400pt}}
\multiput(535.00,126.17)(5.000,-1.000){2}{\rule{1.204pt}{0.400pt}}
\put(515.0,127.0){\rule[-0.200pt]{4.818pt}{0.400pt}}
\put(586,124.67){\rule{2.409pt}{0.400pt}}
\multiput(586.00,125.17)(5.000,-1.000){2}{\rule{1.204pt}{0.400pt}}
\put(545.0,126.0){\rule[-0.200pt]{9.877pt}{0.400pt}}
\put(668,123.67){\rule{2.409pt}{0.400pt}}
\multiput(668.00,124.17)(5.000,-1.000){2}{\rule{1.204pt}{0.400pt}}
\put(596.0,125.0){\rule[-0.200pt]{17.345pt}{0.400pt}}
\put(909,122.67){\rule{2.409pt}{0.400pt}}
\multiput(909.00,123.17)(5.000,-1.000){2}{\rule{1.204pt}{0.400pt}}
\put(678.0,124.0){\rule[-0.200pt]{55.648pt}{0.400pt}}
\put(919.0,123.0){\rule[-0.200pt]{118.041pt}{0.400pt}}
\end{picture}
\caption{Radiatively corrected masses total cross section for the process
$\mu^- \mu^+ \rightarrow H^0 Z^0$ as a function
of $m_{H^0}$. We have taken
$\sqrt{s} = 500 \textrm{GeV/c}$ and $\tan\beta= 30$.}
\label{mumu_HZrc}
\end{center} 
\end{figure}
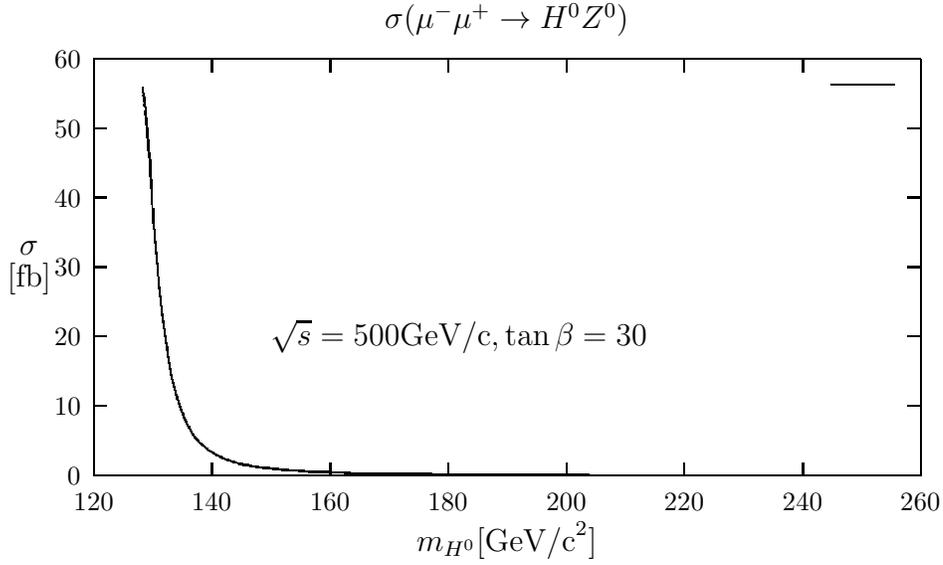

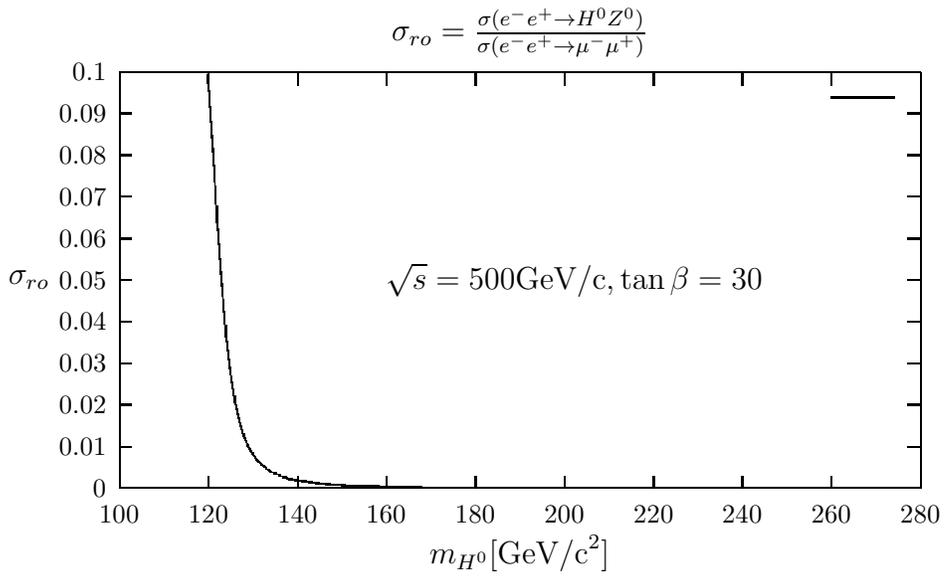
\begin{figure}
\begin{center}
% GNUPLOT: LaTeX picture
\setlength{\unitlength}{0.240900pt}
\ifx\plotpoint\undefined\newsavebox{\plotpoint}\fi
\sbox{\plotpoint}{\rule[-0.200pt]{0.400pt}{0.400pt}}%
\begin{picture}(1500,900)(0,0)
\font\gnuplot=cmr10 at 10pt
\gnuplot
\sbox{\plotpoint}{\rule[-0.200pt]{0.400pt}{0.400pt}}%
\put(181.0,123.0){\rule[-0.200pt]{4.818pt}{0.400pt}}
\put(161,123){\makebox(0,0)[r]{0}}
\put(1419.0,123.0){\rule[-0.200pt]{4.818pt}{0.400pt}}
\put(181.0,188.0){\rule[-0.200pt]{4.818pt}{0.400pt}}
\put(161,188){\makebox(0,0)[r]{0.01}}
\put(1419.0,188.0){\rule[-0.200pt]{4.818pt}{0.400pt}}
\put(181.0,254.0){\rule[-0.200pt]{4.818pt}{0.400pt}}
\put(161,254){\makebox(0,0)[r]{0.02}}
\put(1419.0,254.0){\rule[-0.200pt]{4.818pt}{0.400pt}}
\put(181.0,319.0){\rule[-0.200pt]{4.818pt}{0.400pt}}
\put(161,319){\makebox(0,0)[r]{0.03}}
\put(1419.0,319.0){\rule[-0.200pt]{4.818pt}{0.400pt}}
\put(181.0,385.0){\rule[-0.200pt]{4.818pt}{0.400pt}}
\put(161,385){\makebox(0,0)[r]{0.04}}
\put(1419.0,385.0){\rule[-0.200pt]{4.818pt}{0.400pt}}
\put(181.0,450.0){\rule[-0.200pt]{4.818pt}{0.400pt}}
\put(161,450){\makebox(0,0)[r]{0.05}}
\put(1419.0,450.0){\rule[-0.200pt]{4.818pt}{0.400pt}}
\put(181.0,515.0){\rule[-0.200pt]{4.818pt}{0.400pt}}
\put(161,515){\makebox(0,0)[r]{0.06}}
\put(1419.0,515.0){\rule[-0.200pt]{4.818pt}{0.400pt}}
\put(181.0,581.0){\rule[-0.200pt]{4.818pt}{0.400pt}}
\put(161,581){\makebox(0,0)[r]{0.07}}
\put(1419.0,581.0){\rule[-0.200pt]{4.818pt}{0.400pt}}
\put(181.0,646.0){\rule[-0.200pt]{4.818pt}{0.400pt}}
\put(161,646){\makebox(0,0)[r]{0.08}}
\put(1419.0,646.0){\rule[-0.200pt]{4.818pt}{0.400pt}}
\put(181.0,712.0){\rule[-0.200pt]{4.818pt}{0.400pt}}
\put(161,712){\makebox(0,0)[r]{0.09}}
\put(1419.0,712.0){\rule[-0.200pt]{4.818pt}{0.400pt}}
\put(181.0,777.0){\rule[-0.200pt]{4.818pt}{0.400pt}}
\put(161,777){\makebox(0,0)[r]{0.1}}
\put(1419.0,777.0){\rule[-0.200pt]{4.818pt}{0.400pt}}
\put(181.0,123.0){\rule[-0.200pt]{0.400pt}{4.818pt}}
\put(181,82){\makebox(0,0){100}}
\put(181.0,757.0){\rule[-0.200pt]{0.400pt}{4.818pt}}
\put(321.0,123.0){\rule[-0.200pt]{0.400pt}{4.818pt}}
\put(321,82){\makebox(0,0){120}}
\put(321.0,757.0){\rule[-0.200pt]{0.400pt}{4.818pt}}
\put(461.0,123.0){\rule[-0.200pt]{0.400pt}{4.818pt}}
\put(461,82){\makebox(0,0){140}}
\put(461.0,757.0){\rule[-0.200pt]{0.400pt}{4.818pt}}
\put(600.0,123.0){\rule[-0.200pt]{0.400pt}{4.818pt}}
\put(600,82){\makebox(0,0){160}}
\put(600.0,757.0){\rule[-0.200pt]{0.400pt}{4.818pt}}
\put(740.0,123.0){\rule[-0.200pt]{0.400pt}{4.818pt}}
\put(740,82){\makebox(0,0){180}}
\put(740.0,757.0){\rule[-0.200pt]{0.400pt}{4.818pt}}
\put(880.0,123.0){\rule[-0.200pt]{0.400pt}{4.818pt}}
\put(880,82){\makebox(0,0){200}}
\put(880.0,757.0){\rule[-0.200pt]{0.400pt}{4.818pt}}
\put(1020.0,123.0){\rule[-0.200pt]{0.400pt}{4.818pt}}
\put(1020,82){\makebox(0,0){220}}
\put(1020.0,757.0){\rule[-0.200pt]{0.400pt}{4.818pt}}
\put(1159.0,123.0){\rule[-0.200pt]{0.400pt}{4.818pt}}
\put(1159,82){\makebox(0,0){240}}
\put(1159.0,757.0){\rule[-0.200pt]{0.400pt}{4.818pt}}
\put(1299.0,123.0){\rule[-0.200pt]{0.400pt}{4.818pt}}
\put(1299,82){\makebox(0,0){260}}
\put(1299.0,757.0){\rule[-0.200pt]{0.400pt}{4.818pt}}
\put(1439.0,123.0){\rule[-0.200pt]{0.400pt}{4.818pt}}
\put(1439,82){\makebox(0,0){280}}
\put(1439.0,757.0){\rule[-0.200pt]{0.400pt}{4.818pt}}
\put(181.0,123.0){\rule[-0.200pt]{303.052pt}{0.400pt}}
\put(1439.0,123.0){\rule[-0.200pt]{0.400pt}{157.549pt}}
\put(181.0,777.0){\rule[-0.200pt]{303.052pt}{0.400pt}}
\put(40,450){\makebox(0,0){$\sigma_{ro}$}}
\put(810,21){\makebox(0,0){$m_{H^0} [\textrm{GeV/c}^2]$}}
\put(810,839){\makebox(0,0){$\sigma_{ro}=\frac{\sigma(e^-e^+ \rightarrow H^0 Z^0)}{\sigma(e^- e^+ \rightarrow \mu^- \mu^+)}$}}
\put(600,450){\makebox(0,0)[l]{$\sqrt{s} = 500 \textrm{GeV/c}, \tan\beta = 
30$}}
\put(181.0,123.0){\rule[-0.200pt]{0.400pt}{157.549pt}}
\put(1279,737){\makebox(0,0)[r]{ }}
\put(1299.0,737.0){\rule[-0.200pt]{24.090pt}{0.400pt}}
\put(321,772){\usebox{\plotpoint}}
\multiput(321.59,746.20)(0.485,-8.120){11}{\rule{0.117pt}{6.214pt}}
\multiput(320.17,759.10)(7.000,-94.102){2}{\rule{0.400pt}{3.107pt}}
\multiput(328.59,638.02)(0.485,-8.502){11}{\rule{0.117pt}{6.500pt}}
\multiput(327.17,651.51)(7.000,-98.509){2}{\rule{0.400pt}{3.250pt}}
\multiput(335.59,528.39)(0.485,-7.739){11}{\rule{0.117pt}{5.929pt}}
\multiput(334.17,540.69)(7.000,-89.695){2}{\rule{0.400pt}{2.964pt}}
\multiput(342.59,430.90)(0.485,-6.290){11}{\rule{0.117pt}{4.843pt}}
\multiput(341.17,440.95)(7.000,-72.948){2}{\rule{0.400pt}{2.421pt}}
\multiput(349.59,352.88)(0.485,-4.688){11}{\rule{0.117pt}{3.643pt}}
\multiput(348.17,360.44)(7.000,-54.439){2}{\rule{0.400pt}{1.821pt}}
\multiput(356.59,295.15)(0.485,-3.315){11}{\rule{0.117pt}{2.614pt}}
\multiput(355.17,300.57)(7.000,-38.574){2}{\rule{0.400pt}{1.307pt}}
\multiput(363.59,253.76)(0.485,-2.476){11}{\rule{0.117pt}{1.986pt}}
\multiput(362.17,257.88)(7.000,-28.879){2}{\rule{0.400pt}{0.993pt}}
\multiput(370.59,223.13)(0.485,-1.713){11}{\rule{0.117pt}{1.414pt}}
\multiput(369.17,226.06)(7.000,-20.065){2}{\rule{0.400pt}{0.707pt}}
\multiput(377.59,201.55)(0.485,-1.255){11}{\rule{0.117pt}{1.071pt}}
\multiput(376.17,203.78)(7.000,-14.776){2}{\rule{0.400pt}{0.536pt}}
\multiput(384.59,185.74)(0.485,-0.874){11}{\rule{0.117pt}{0.786pt}}
\multiput(383.17,187.37)(7.000,-10.369){2}{\rule{0.400pt}{0.393pt}}
\multiput(391.59,174.21)(0.485,-0.721){11}{\rule{0.117pt}{0.671pt}}
\multiput(390.17,175.61)(7.000,-8.606){2}{\rule{0.400pt}{0.336pt}}
\multiput(398.00,165.93)(0.492,-0.485){11}{\rule{0.500pt}{0.117pt}}
\multiput(398.00,166.17)(5.962,-7.000){2}{\rule{0.250pt}{0.400pt}}
\multiput(405.00,158.93)(0.581,-0.482){9}{\rule{0.567pt}{0.116pt}}
\multiput(405.00,159.17)(5.824,-6.000){2}{\rule{0.283pt}{0.400pt}}
\multiput(412.00,152.93)(0.710,-0.477){7}{\rule{0.660pt}{0.115pt}}
\multiput(412.00,153.17)(5.630,-5.000){2}{\rule{0.330pt}{0.400pt}}
\multiput(419.00,147.95)(1.355,-0.447){3}{\rule{1.033pt}{0.108pt}}
\multiput(419.00,148.17)(4.855,-3.000){2}{\rule{0.517pt}{0.400pt}}
\multiput(426.00,144.95)(1.355,-0.447){3}{\rule{1.033pt}{0.108pt}}
\multiput(426.00,145.17)(4.855,-3.000){2}{\rule{0.517pt}{0.400pt}}
\multiput(433.00,141.95)(1.355,-0.447){3}{\rule{1.033pt}{0.108pt}}
\multiput(433.00,142.17)(4.855,-3.000){2}{\rule{0.517pt}{0.400pt}}
\put(440,138.17){\rule{1.500pt}{0.400pt}}
\multiput(440.00,139.17)(3.887,-2.000){2}{\rule{0.750pt}{0.400pt}}
\put(447,136.17){\rule{1.500pt}{0.400pt}}
\multiput(447.00,137.17)(3.887,-2.000){2}{\rule{0.750pt}{0.400pt}}
\put(454,134.67){\rule{1.686pt}{0.400pt}}
\multiput(454.00,135.17)(3.500,-1.000){2}{\rule{0.843pt}{0.400pt}}
\put(461,133.67){\rule{1.686pt}{0.400pt}}
\multiput(461.00,134.17)(3.500,-1.000){2}{\rule{0.843pt}{0.400pt}}
\put(468,132.67){\rule{1.686pt}{0.400pt}}
\multiput(468.00,133.17)(3.500,-1.000){2}{\rule{0.843pt}{0.400pt}}
\put(475,131.67){\rule{1.686pt}{0.400pt}}
\multiput(475.00,132.17)(3.500,-1.000){2}{\rule{0.843pt}{0.400pt}}
\put(482,130.67){\rule{1.686pt}{0.400pt}}
\multiput(482.00,131.17)(3.500,-1.000){2}{\rule{0.843pt}{0.400pt}}
\put(489,129.67){\rule{1.686pt}{0.400pt}}
\multiput(489.00,130.17)(3.500,-1.000){2}{\rule{0.843pt}{0.400pt}}
\put(502,128.67){\rule{1.686pt}{0.400pt}}
\multiput(502.00,129.17)(3.500,-1.000){2}{\rule{0.843pt}{0.400pt}}
\put(509,127.67){\rule{1.686pt}{0.400pt}}
\multiput(509.00,128.17)(3.500,-1.000){2}{\rule{0.843pt}{0.400pt}}
\put(496.0,130.0){\rule[-0.200pt]{1.445pt}{0.400pt}}
\put(530,126.67){\rule{1.686pt}{0.400pt}}
\multiput(530.00,127.17)(3.500,-1.000){2}{\rule{0.843pt}{0.400pt}}
\put(516.0,128.0){\rule[-0.200pt]{3.373pt}{0.400pt}}
\put(551,125.67){\rule{1.686pt}{0.400pt}}
\multiput(551.00,126.17)(3.500,-1.000){2}{\rule{0.843pt}{0.400pt}}
\put(537.0,127.0){\rule[-0.200pt]{3.373pt}{0.400pt}}
\put(586,124.67){\rule{1.686pt}{0.400pt}}
\multiput(586.00,125.17)(3.500,-1.000){2}{\rule{0.843pt}{0.400pt}}
\put(558.0,126.0){\rule[-0.200pt]{6.745pt}{0.400pt}}
\put(649,123.67){\rule{1.686pt}{0.400pt}}
\multiput(649.00,124.17)(3.500,-1.000){2}{\rule{0.843pt}{0.400pt}}
\put(593.0,125.0){\rule[-0.200pt]{13.490pt}{0.400pt}}
\put(817,122.67){\rule{1.686pt}{0.400pt}}
\multiput(817.00,123.17)(3.500,-1.000){2}{\rule{0.843pt}{0.400pt}}
\put(656.0,124.0){\rule[-0.200pt]{38.785pt}{0.400pt}}
\put(824.0,123.0){\rule[-0.200pt]{129.604pt}{0.400pt}}
\end{picture}
\caption{Total cross section for the process
$e^- e^+ \rightarrow H^0 Z^0$ compared with the
cross section $\sigma(e^- e^+
\rightarrow \mu^- \mu^+)$ as a function
of $m_{H^0}$. We have taken
$\sqrt{s} = \textrm{500 GeV/c}$ and $\tan\beta= 30$. The radiative 
corrections of the masses
were not taken into account.}
\label{muhzee_figure}  
\end{center}
\end{figure}

\begin{figure} 
\begin{center}
\input{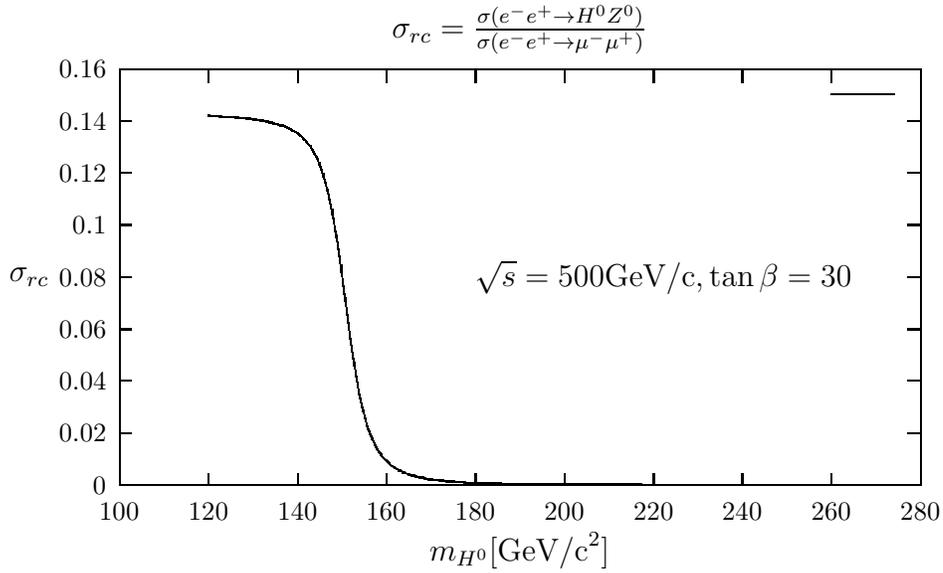}
\caption{Radiatively corrected masses total cross section for the process
$e^- e^+ \rightarrow H^0 Z^0$ compared with the
cross section $\sigma(e^- e^+
\rightarrow \mu^- \mu^+)$ as a function
of $m_{H^0}$. We have taken
$\sqrt{s} = 500 \textrm{GeV/c}$ and $\tan\beta= 30$.}
\label{muhzeerc_figure}
\end{center}   
\end{figure}

The total cross section corresponding to 
$\mu^- \mu^+ \rightarrow H^0 Z^0$ is obtained from Equation
(\ref{sigma_muhz}) replacing $\left(\tan\beta - \tan\alpha
\right)^2$ by $\left( 1 + \tan\beta \tan\alpha \right)^2$
in the numerator and $m_{h^0}$ by $m_{H^0}$. This production 
cross section is plotted in Figures \ref{mumu_HZ1},
\ref{mumu_HZrc} as a 
function of $m_{H^0}$ for $\sqrt{s} = 500 \textrm{GeV/c}$ and 
$\tan\beta = 30$, without and with mass radiative corrections, 
respectively.
In Figure \ref{muhzee_figure} we show the ratio
between the production cross section
$\sigma \left(e^- e^+ \rightarrow H^0 Z^0 \right)$
and the cross section 
$\sigma \left(e^- e^+ \rightarrow \mu^- \mu^+ \right)$ in terms
of $m_{H^0}$. The radiatively corrected masses total cross section is 
shown 
in Figure \ref{muhzeerc_figure}. Figures \ref{mumu_HZrc} and
\ref{muhzeerc_figure} show the importance of the radiative corrections
of the masses
in the processes $\mu^- \mu^+ \rightarrow H^0 Z^0$ and $e^- e^+ 
\rightarrow H^0 Z^0$. 

\section{Production of $A^0$}
From the Feynman diagrams of Figure \ref{mumu_AZ_fig} and the Feynman 
rules given in \cite{M_H}, we obtain the
differential cross section for the production process
$\mu^- \mu^+ \rightarrow A^0 Z^0$ in the center of mass system:
\begin{figure}  
\begin{center}   
%\vspace*{-3.5cm}
%\scalebox{0.5}
{\includegraphics{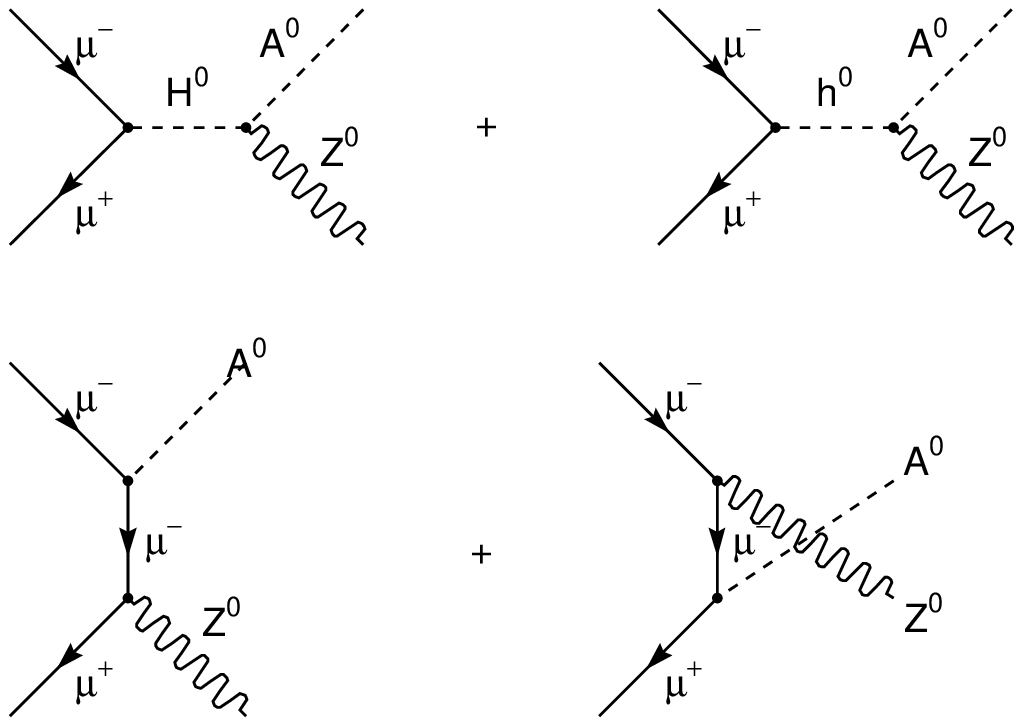}}
%\vspace*{0.7cm}
\caption{Feynman diagrams corresponding to the production
of $A^0$ in the channel $\mu^- \mu^+ \rightarrow A^0 Z^0$.}
\label{mumu_AZ_fig}
\end{center}
\end{figure}  

\begin{eqnarray}
\lefteqn{
\frac{d \sigma}{d \Omega} (\mu^- \mu^+ \rightarrow A^0 Z^0) =
\frac{1}{64 \pi^2 s} \Lambda^{1/2}\left( s, m_{A^0}^2, 
m_Z^2 \right) G_F^2 m_{\mu}^2 
}
\nonumber \\ & &
\{  C_{Hb}^2 
\Lambda \left( s, m_{A^0}^2, m_Z^2 \right)
+ \left[  \left( g_A^\mu \right)^2 + \left( g_V^\mu \right)^2
\right] 
\nonumber \\ & &
[ \tan^2\beta \left( 1 +
 \frac{\Lambda \left( s, m_{A^0}^2, m_Z^2 \right)m_Z^2
\sin^2\theta}{2s t^2}\right)
\nonumber \\ & &
 + \tan^2\beta 
\left( 1 +  \frac{\Lambda \left( s, m_{A^0}^2, m_Z^2 \right)m_Z^2
\sin^2\theta}{2s u^2}\right) ]
\nonumber \\ & &
+ 2 g_A^\mu \tan\beta C_{Hb}
\left[\frac{m_{A^0}^2 m_Z^2}{t} - 
\frac{\Lambda \left( s, m_{A^0}^2, m_Z^2 \right) \sin^2\theta}
{4t} - t \right]
\nonumber \\ & &
+ 2 g_A^\mu \tan\beta C_{Hb}
\left[\frac{m_{A^0}^2 m_Z^2}{u} -      
\frac{\Lambda \left( s, m_{A^0}^2, m_Z^2 \right) \sin^2\theta}
{4u} - u \right] 
\nonumber \\ & &
+ 2\tan^2\beta
 \left[  \left( g_V^\mu \right)^2 - \left( g_A^\mu \right)^2
\right]
[\frac{m_{A^0}^2 m_Z^2}{ut} -
\frac{\Lambda \left( s, m_{A^0}^2, m_Z^2 \right) \sin^2\theta}
{4ut}
\nonumber \\ & &
 + \frac{ \Lambda
 \left( s, m_{A^0}^2, m_Z^2 \right) m_Z^2 \sin^2\theta }
{2sut} ] \}
\label{diffmumu_AZ}
\end{eqnarray}
\noindent
where $g_A^\mu$ and $g_V^\mu$ are given by 
(\ref{gA,gV}); $s$,$t$,$u$ are the
Mandelstam invariant variables and

\begin{equation}
C_{Hb} = \frac{ \left( \frac{1}{2} \sin 2 \alpha +
\tan\beta \sin^2\alpha \right) }{\left (s - m_{h^0}^2
\right) }
- \frac{ \left( \frac{1}{2} \sin2\alpha - \tan\beta
\cos^2\alpha \right) }{ \left( s - m_{H^0}^2 \right)}
\label{CHb}
\end{equation}

To obtain the total cross section, we integrate
Equation (\ref{diffmumu_AZ}) over the solid angle $\Omega$.
\begin{eqnarray}
\lefteqn{
\sigma  (\mu^- \mu^+ \rightarrow A^0 Z^0) = 
\frac{G_F^2 m_{\mu}^2}{16 \pi s^2}
\{\Lambda^{1/2} \left( s, m_{A^0}^2, m_Z^2 \right)
[ s C_{Hb}^2 \Lambda \left( s, m_{A^0}^2, m_Z^2 \right)
}
\nonumber \\ & &
+ 4 \tan^2\beta \sin^2\theta_W \left(1 -2 \sin^2\theta_W
\right) \left( s - 2 m_Z^2 \right) + 2 s \tan\beta  C_{Hb}
\nonumber \\ & &
\times \left( m_{A^0}^2 + m_Z^2 - s \right)
+ \left( 1 - 4 \sin^2\theta_W + 8 \sin^4\theta_W \right)
\nonumber \\ & &
\times\tan^2\beta \left( s - 4 m_Z^2 \right) ]
+ 4 m_Z^2 \tan\beta f\left( s, m_{A^0}^2, m_Z^2 \right)
\nonumber \\ & &
\times[ - s C_{Hb} m_{A^0}^2 + \frac{1}{2} \tan\beta \left(1 - 
4 \sin^2\theta_W + 8 \sin^4\theta_W \right)
\left( m_{A^0}^2 + m_Z^2 - s \right)
\nonumber \\ & &
- 4 \frac{\sin^2\theta_W \left( 1 - 2 \sin^2\theta_W\right)
\tan\beta m_{A^0}^2 \left( s - m_Z^2 \right)}
{\left( m_{A^0}^2 + m_Z^2 - s \right)}]\}
\nonumber \\ & &
\times
\left( 3.8938 \times 10^{11} \right) \textrm{fb}
\label{sigmamumu_AZ}
\end{eqnarray}

\noindent where

\begin{equation}
f \left( s, m_{A^0}^2, m_Z^2 \right) \equiv
\ln\left| \frac{m_{A^0}^2 + m_Z^2 - s +
\Lambda^{1/2} \left( s, m_{A^0}^2, m_Z^2 \right)}
{ m_{A^0}^2 + m_Z^2 - s -
\Lambda^{1/2} \left( s, m_{A^0}^2, m_Z^2 \right)}
\right|
\label{f}
\end{equation}

Note that if $m_{A^0} = \sqrt{s} - m_Z$, then we have,
$\Lambda \left( s, m_{A^0}^2, m_Z^2 \right) = 0$ and
$f \left( s, m_{A^0}^2, m_Z^2 \right)$ = 0. Therefore
$\sigma  (\mu^- \mu^+ \rightarrow A^0 Z^0) =0$.

\begin{figure}
\begin{center}
\input{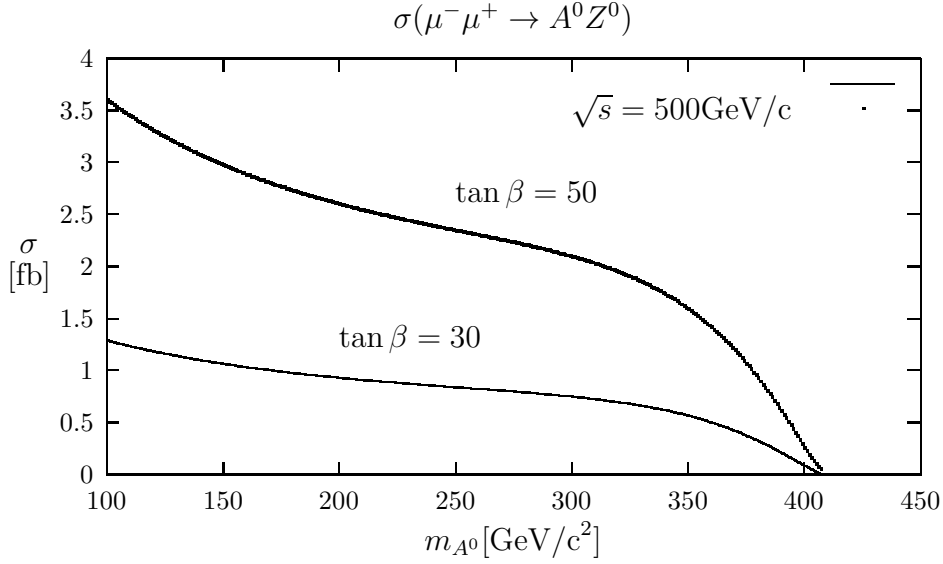}
\caption{Total cross section for the process
$\mu^- \mu^+ \rightarrow A^0 Z^0$ as a function
of $m_{A^0}$. We have taken
$\sqrt{s} = 500 \textrm{GeV/c}$ and $\tan\beta= 30, 50$. The total 
cross section is not affected by radiative corrections of the masses.}
\label{muaz_figure}
\end{center}
\end{figure}

Figure \ref{muaz_figure} shows the total 
cross section $\sigma  (\mu^- \mu^+ \rightarrow A^0 Z^0)$
as a function of $m_{A^0}$ for $\sqrt{s} = 500 \textrm{GeV/c}$ and 
$\tan\beta= 30, 50$. The total cross section is not
affected by radiative corrections of the masses. From Figure 
\ref{muaz_figure} we can 
see that cross sections are important for large values of $\tan\beta$.

The total cross section corresponding to 
$e^- e^+ \rightarrow A^0 Z^0$ can be obtained from Equation
(\ref{sigmamumu_AZ}) replacing $m_{\mu}$ by $m_{e}$:
\begin{equation}
\frac{\sigma(e^- e^+ \rightarrow A^0 Z^0)}
{\sigma(\mu^- \mu^+ \rightarrow A^0 Z^0)} = 
\left(\frac{m_{e}}{m_{\mu}}\right)^2 = 2.34 \cdot 10^{-5}.
\label{sigmaeeAZ}
\end{equation}
 
\section{Production of $H^{\pm}$}

\begin{figure}
\begin{center}
%\vspace*{-4.5cm}
%\scalebox{0.5}
{\includegraphics{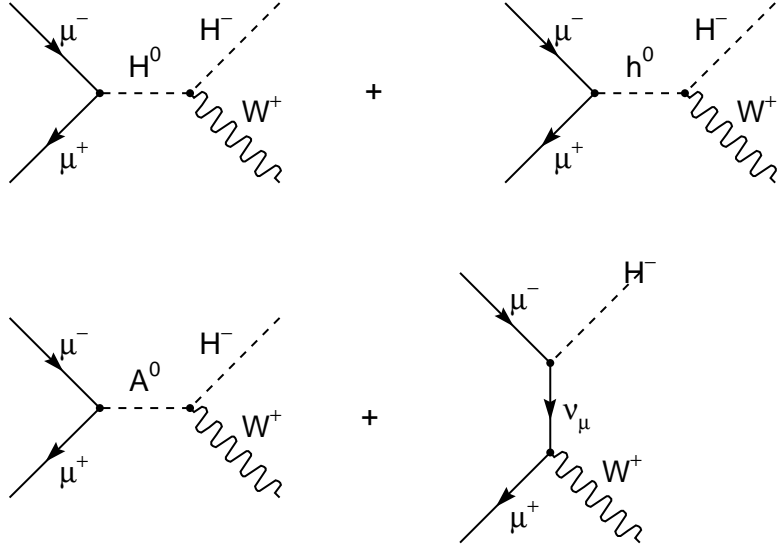}}
%\vspace*{0.7cm}
\caption{Feynman diagrams corresponding to the production
of $H^-$ in the channel $\mu^- \mu^+ \rightarrow H^-
W^+$.}
\label{mumu_HW_fig}
\end{center}
\end{figure}

From the Feynman diagrams of Figure \ref{mumu_HW_fig}
we obtain the differential cross section
in the center of mass system for the process
$\mu^- \mu^+ \rightarrow H^- W^+$:

\begin{eqnarray}
\lefteqn{
\frac{d \sigma}{d \Omega} (\mu^- \mu^+ \rightarrow H^- W^+) =
\frac{1}{64 \pi^2 s} \Lambda^{1/2}\left( s, m_H^2,
m_W^2 \right) G_F^2 m_{\mu}^2
}
\nonumber \\ & &
\times\{ \left[ C_{Hb}^2 + C_{Ab}^2 \right] \Lambda\left( s, m_H^2, 
m_W^2 \right) 
+ 2 \left( \frac{\tan\beta}{t} \right)^2
\nonumber \\ & &
\times\left[ \frac{\Lambda\left( s, m_H^2, m_W^2 \right)
\sin^2\theta m_W^2}{2s} + t^2 \right]
- 2  \left( \frac{\tan\beta}{t} \right) \left(C_{Ab} + C_{Hb} 
\right)
\nonumber \\ & &
\times\left[ -t^2 - \frac{1}{4} 
\Lambda\left( s, m_H^2, m_W^2 \right) \sin^2\theta
+ m_W^2 m_H^2 \right] \}
\label{diffmumu_HW}
\end{eqnarray}

\noindent where $C_{Hb}$ is given by Equation (\ref{CHb}) and 
\begin{equation}
C_{Ab} = \frac{\tan\beta}{\left( s - m_{A^0}^2 \right)}
\label{CAb}
\end{equation}

The differential cross section corresponding to
$\mu^- \mu^+ \rightarrow H^+
W^-$ is obtained from  (\ref{diffmumu_HW}) by replacing $t$ by
$u$.

The integration of (\ref{diffmumu_HW})
over the solid angle $\Omega$ give us the total cross
section:

\begin{eqnarray}
\lefteqn{
\sigma (\mu^- \mu^+ \rightarrow H^- W^+) =
\frac{G_F^2 m_{\mu}^2}{16 \pi s^2} \{ s 
\Lambda^{3/2}\left( s, m_H^2,m_W^2 \right)
\left[ C_{Hb}^2 + C_{Ab}^2 \right]
}
\nonumber \\ & &
+ 2 \tan\beta \Lambda^{1/2}\left( s, m_H^2,m_W^2 \right)
[ \tan\beta \left( s - 4 m_W^2 \right) 
\nonumber \\ & &
+ \left( C_{Ab} +
C_{Hb} \right) s \left( m_H^2 + m_W^2 - s \right)]
+ 4 m_W^2 \tan\beta f \left( s, m_H^2, m_W^2 \right)
\nonumber \\ & &
\left[ \tan\beta \left( m_H^2 + m_W^2 - s \right)
- \left( C_{Ab} + C_{Hb} \right) s m_H^2 \right] \}
\nonumber \\ & &
\times
\left( 3.8938 \times 10^{11} \right) \textrm{fb}
\label{sigmamumu_HW}
\end{eqnarray}

\noindent where 

\begin{equation}
f \left( s, m_H^2, m_W^2 \right) =
\ln\left| \frac{m_H^2 + m_W^2 - s +
\Lambda^{1/2} \left( s, m_H^2, m_W^2 \right)}
{ m_H^2 + m_W^2 - s -
\Lambda^{1/2} \left( s, m_H^2, m_W^2 \right)}
\right|
\label{fW}
\end{equation}

For the process $ \mu^- \mu^+ \rightarrow H^+ W^-$ we obtain:

\begin{equation}
\sigma (\mu^- \mu^+ \rightarrow H^+ W^-) =
\sigma (\mu^- \mu^+ \rightarrow H^- W^+)
\label{H+/-W-/+}
\end{equation}

\noindent and then

\begin{equation}
\sigma (\mu^- \mu^+ \rightarrow H^{\pm} W^{\mp}) = 2
\sigma (\mu^- \mu^+ \rightarrow H^- W^+)
\label{sigmaH+/-W-/+}
\end{equation}

Observe that $\sigma (\mu^- \mu^+ \rightarrow H^{\pm} W^{\mp})
= 0$ if $m_H = \sqrt{s} - m_W$.

\begin{figure}
\begin{center}
\input{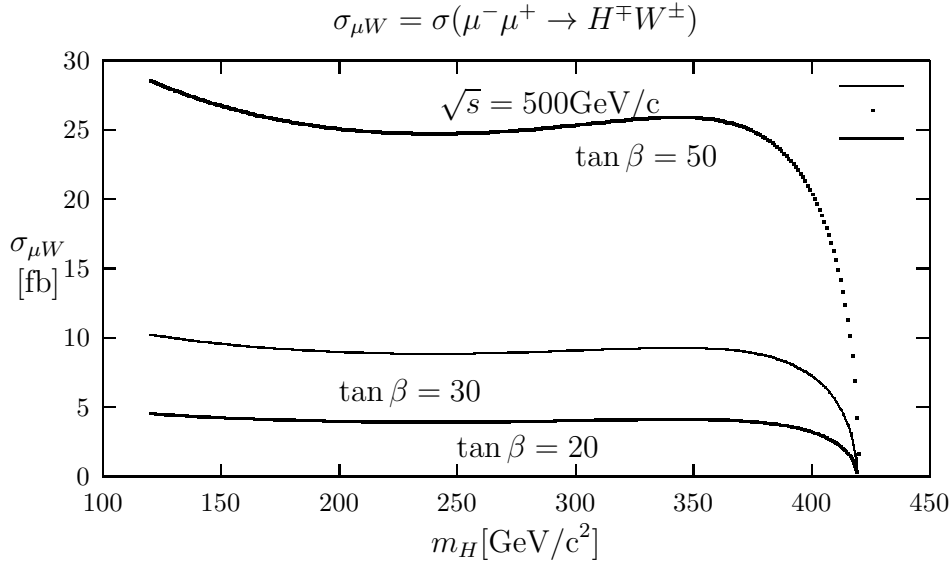}
\caption{Total cross section for the process
$\mu^- \mu^+ \rightarrow H^{\mp} W^{\pm}$ as a function
of $m_H$. We have taken
$\sqrt{s} = 500 \textrm{GeV/c}$ and $\tan\beta= 20, 30, 50$. The radiative 
corrections of the masses
are negligible.}
\label{muhw_figure}
\end{center}
\end{figure}

The total cross section corresponding to $\mu^- \mu^+ 
\rightarrow H^{\mp} W^{\pm}$ is given in Figure \ref{muhw_figure} for 
$\sqrt{s} = 500 \textrm{GeV/c}$ and $\tan\beta = 20, 30 , 50$. 
This total cross
section is not affected by radiative corrections of the masses. From 
Figure 
\ref{muhw_figure} we see that 
$\sigma(\mu^- \mu^+ \rightarrow H^{\mp} W^{\pm}) \gtrsim 5 \textrm{fb}$ 
for 
$\tan\beta \geq 20$ in the mass interval
$100 \leq m_{H} \leq 400 [\textrm{GeV/c}^2]$.

For the process
$e^- e^+ \rightarrow H^{\mp} W^{\pm}$, the total cross section 
is obtained from Equations (\ref{sigmamumu_HW}), (\ref{sigmaH+/-W-/+})  
replacing $m_{\mu}$ by $m_{e}$. This cross section is smaller than the
one ploted in Figure \ref{muhw_figure} by a factor  
$ m_{e}^2/m_{\mu}^2 = 2.34\cdot10^{-5}$. 

\section{Production of charged Higgs boson pairs}

\begin{figure}
\begin{center}
%\vspace*{-4.5cm}
%\scalebox{0.5}
{\includegraphics{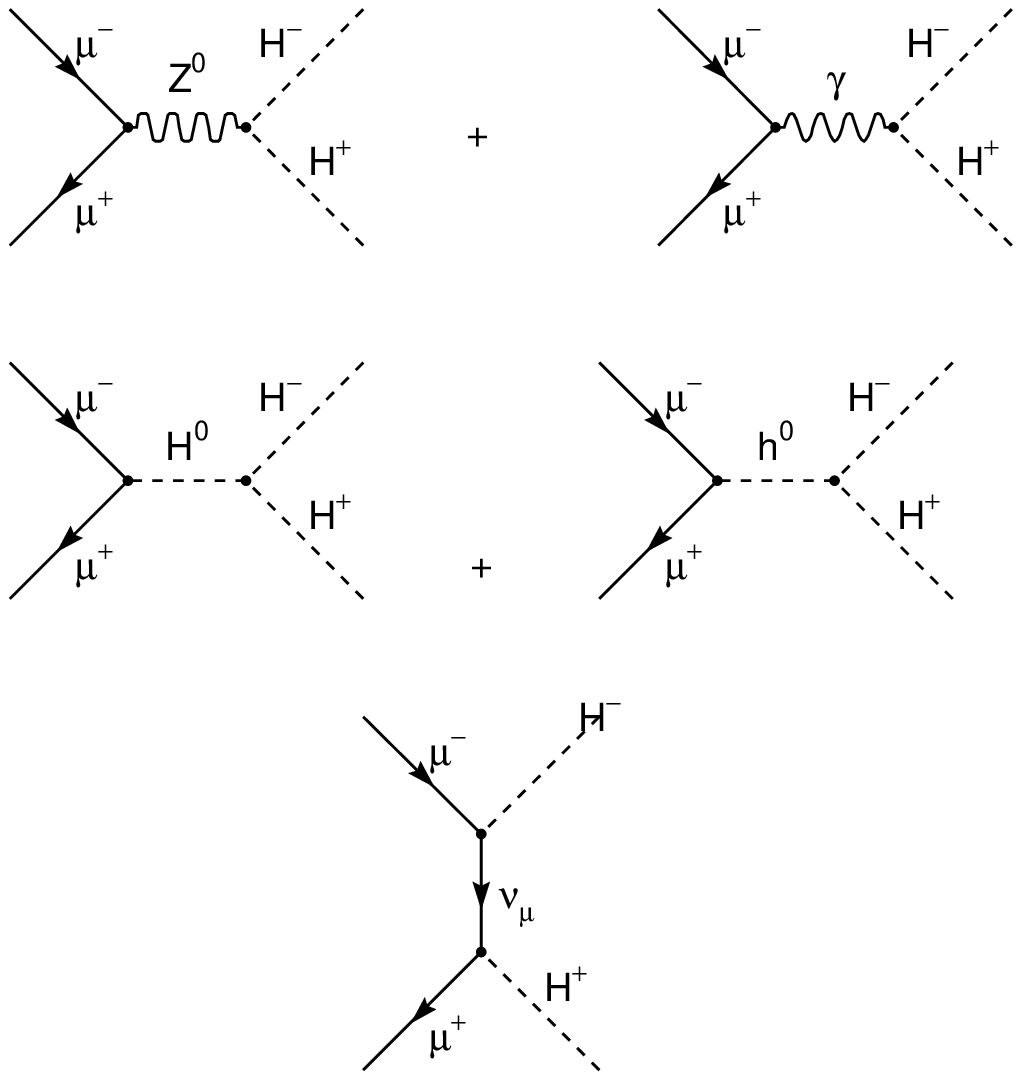}}
%\vspace*{0.7cm}
\caption{Feynman diagrams corresponding to the production
of charged higgs boson pairs
in the channel $\mu^- \mu^+ \rightarrow H^-
H^+$.}
\label{mumu_HH_fig}
\end{center}
\end{figure}

From the Feynman diagrams of Figure \ref{mumu_HH_fig}, the differential 
cross section in the center of mass system corresponding to
$\mu^- \mu^+ \rightarrow H^- H^+$ is

\begin{eqnarray}
\lefteqn{
\frac{d \sigma}{d \Omega} (\mu^- \mu^+ \rightarrow H^- H^+) =
\frac{G_F^2 m_W^4}{8 \pi^2 s} \left( 1 - 4 \frac{m_H^2}{s}
\right)^{1/2} \{ s \left(s - 4m_H^2 \right) \sin^2\theta
}
\nonumber \\ & &
\left[ \frac{1}{8} \left|C_1 \right|^2\left[ 
\left( g_A^\mu \right)^2 + \left( g_V^\mu \right)^2
\right] + 2 \left( \frac{\sin^2\theta_W}{s} \right)^2
- \left( \frac{\sin^2\theta_W}{s} \right) \Re{(C_1)}
g_V^{\mu} \right] 
\nonumber \\ & &
+ 2 m_{\mu}^2 [ \left(s - 4 m_H^2 
\right) 
[ \frac{\left| C_1 \right|^2}{4}
\left( \cos^2\theta \left( g_V^{\mu} \right)^2
-\sin^2\theta \left( g_A^{\mu} \right)^2 \right)
\nonumber \\ & &
+ 4 \left( \frac{\sin^2\theta_W}{s} \right)^2
\cos^2\theta
- 2  \left( \frac{\sin^2\theta_W}{s} \right)
\Re{(C_1)} g_V^{\mu} \cos^2\theta ]
\nonumber \\ & &
+ \frac{1}{4} \left( C^{Hh} \right)^2 s 
+ \left( s \left( s - 4 m_H^2 \right) \right)^{1/2}
\nonumber \\ & &
\times\cos\theta C^{Hh} \left( \frac{1}{2} g_V^{\mu}
\Re{(C_1)} - 2 \left( \frac{\sin^2\theta_W}{s} \right)
\right) ] \}
\label{diffmumu_HH}
\end{eqnarray}

\noindent where $g_A^{\mu}$ and $g_V^{\mu}$ are given by Equation
(\ref{gA,gV}),

\begin{equation}
C_1 \equiv \frac{\cos(2\theta_W)}{\cos^2\theta_W}
\frac{1}{\left( s - m_Z^2 + i m_Z \Gamma_Z \right)},
\label{C1}
\end{equation}

\begin{equation}
C^{Hh} = \frac{a_1}{\left(s - m_{H^0}^2 \right)}
- \frac{a_2}{\left( s - m_{h^0}^2 \right)},
\label{CHh}
\end{equation}

\begin{eqnarray}
\lefteqn{
a_1 = [ \cos^2\alpha + \frac{\tan\beta \sin2\alpha}{2}
- \frac{m_Z}{2m_W \cos\theta_W}
}
\nonumber \\ & &
\frac{\left( 1 - \tan^2\beta \right)}
{\left( 1 + \tan^2\beta \right)}
\left( \cos^2\alpha - \frac{\tan\beta \sin2\alpha}{2}
\right) ],
\label{a1}
\end{eqnarray}

\begin{eqnarray}
\lefteqn{
a_2 = [ \frac{\tan\beta \sin2\alpha}{2}
-\sin^2\alpha + \frac{m_Z}{2m_W \cos\theta_W}
}
\nonumber \\ & &
\frac{\left( 1 - \tan^2\beta \right)}
{\left( 1 + \tan^2\beta \right)}
\left( \sin^2\alpha + \frac{\tan\beta \sin2\alpha}{2}
\right) ]
\label{a2}
\end{eqnarray}

The integration of (\ref{diffmumu_HH}) give us the total 
cross section for the process $\mu^- \mu^+ \rightarrow H^- H^+$:

\begin{eqnarray}
\lefteqn{
\sigma\left(\mu^+ \mu^- \rightarrow H^+ H^- \right) =
\sigma\left(\mu^- \mu^+ \rightarrow H^- H^+ \right) =
\frac{2m_W^4 G_F^2 \sin^4\theta_W}{3\pi s}
}
\nonumber \\ & &
\left( 1 - \frac{4 m_H^2}{s} \right)^{3/2}
\{ [ 1 + \frac{\left(1 -2 \sin^2\theta_W \right)^2
\left(1 + \left(4\sin^2\theta_W -1 \right)^2
\right)  }
{64 \sin^4\theta_W \cos^4\theta_W}
\nonumber \\ & &
\times\frac{1}{\left[ \left( 1 - \frac{m_Z^2}{s}\right)^2
+ \left(\frac{M_Z\Gamma_Z}{s}\right)^2 \right]}
- \frac{\left(1 -2 \sin^2\theta_W\right)
\left(4\sin^2\theta_W -1\right)}
{4 \sin^2\theta_W\cos^2\theta_W}
\nonumber \\ & &
\times\frac{\left(1- \frac{m_Z^2}{s} \right)}
{\left[ \left( 1 - \frac{m_Z^2}{s}\right)^2  
+ \left(\frac{M_Z\Gamma_Z}{s}\right)^2 \right]} ]
+ \frac{m_{\mu}^2}{s} [ \frac{\left(\left(4
\sin^2\theta_W -1\right)^2-2\right)}
{32\sin^4\theta_W\cos^4\theta_W}
\nonumber \\ & &
\times\frac{\left(1 - 2\sin^2\theta_W
\right)^2}{\left[\left( 1 - \frac{m_Z^2}{s}\right)^2
+ \left(\frac{M_Z\Gamma_Z}{s}\right)^2 \right]}
+2 - \frac{\left( 1 - 2 \sin^2\theta_W\right)
\left(4 \sin^2\theta_W -1\right)}
{2\sin^2\theta_W\cos^2\theta_W}
\nonumber \\ & &
\times\frac{\left(1 - \frac{m_Z^2}{s}\right)}
{\left[\left( 1 - \frac{m_Z^2}{s}\right)^2
+ \left(\frac{M_Z\Gamma_Z}{s}\right)^2 \right]}
+ \frac{3}{4}\frac{s^2 (C^{Hh})^2}
{\left(1-4\frac{m_H^2}{s}\right)} ] \}
\nonumber \\ & &
\times
\left( 3.8938 \times 10^{11} \right) \textrm{fb}
\label{sigmamumu_HH}
\end{eqnarray}

Neglecting the mass of the muon we can write:
\begin{eqnarray}
\lefteqn{
\sigma\left(\mu^+ \mu^- \rightarrow H^+ H^- \right) =
\sigma\left(\mu^- \mu^+ \rightarrow H^- H^+ \right) =
\frac{2m_W^4 G_F^2 \sin^4\theta_W}{3\pi s}
}
\nonumber \\ & &
\left( 1 - \frac{4 m_H^2}{s} \right)^{3/2}
\{  1 + \frac{\left(1 -2 \sin^2\theta_W \right)^2
\left(1 + \left(4\sin^2\theta_W -1 \right)^2 
\right)  }
{64 \sin^4\theta_W \cos^4\theta_W}
\nonumber \\ & &
\times\frac{1}{\left[ \left( 1 - \frac{m_Z^2}{s}\right)^2
+ \left(\frac{M_Z\Gamma_Z}{s}\right)^2 \right]}
- \frac{\left(1 -2 \sin^2\theta_W\right)
\left(4\sin^2\theta_W -1\right)}
{4 \sin^2\theta_W\cos^2\theta_W}
\nonumber \\ & &
\times\frac{\left(1- \frac{m_Z^2}{s} \right)}
{\left[ \left( 1 - \frac{m_Z^2}{s}\right)^2
+ \left(\frac{M_Z\Gamma_Z}{s}\right)^2 \right]} \}
\times
\left( 3.8938 \times 10^{11} \right) \textrm{fb}
\label{sigmareducedmumu_HH}
\end{eqnarray}

\begin{figure}
\begin{center}
\input{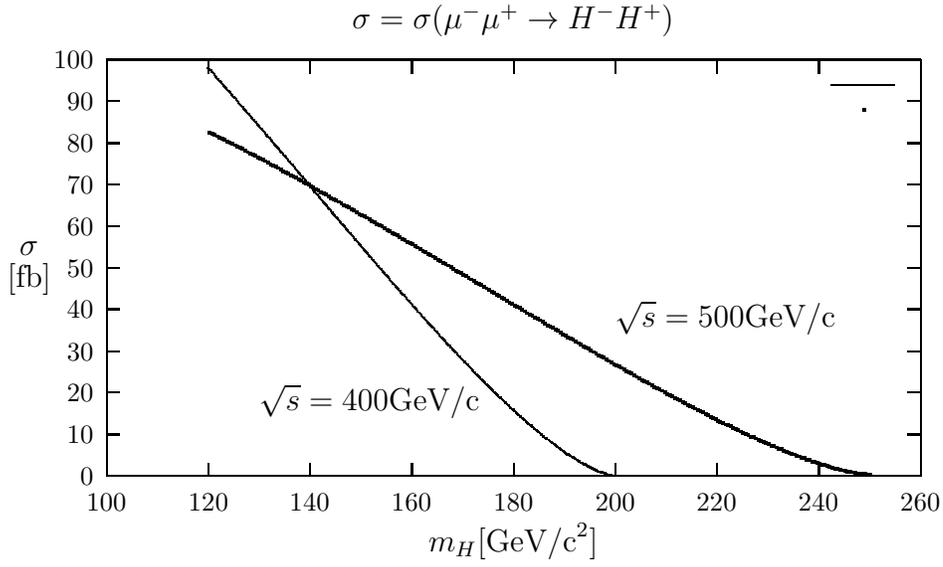}
\caption{Total cross section for the process
$\mu^- \mu^+ \rightarrow H^- H^+$ as a function
of $m_H$. We have taken
$\sqrt{s} = 400, 500 \textrm{GeV/c}$. The total cross section is
practically independent of $\tan\beta$. The radiative 
corrections of the masses
are negligible.}
\label{sigmamumu->HH_fig}
\end{center}
\end{figure}

In the last approximation there is no difference with the total cross
section corresponding to the process $e^- e^+ \rightarrow H^- H^+$.
In Figure \ref{sigmamumu->HH_fig} we have plotted 
the total cross section given by Equation
(\ref{sigmamumu_HH}) as a function of the mass of the
charged higgs for $\sqrt{s}= 400, 500 \textrm{GeV/c}$. The total
cross section is practically independent of $\tan\beta$.
The radiative corrections of the masses are also negligible. 

\begin{figure}
\begin{center}
\input{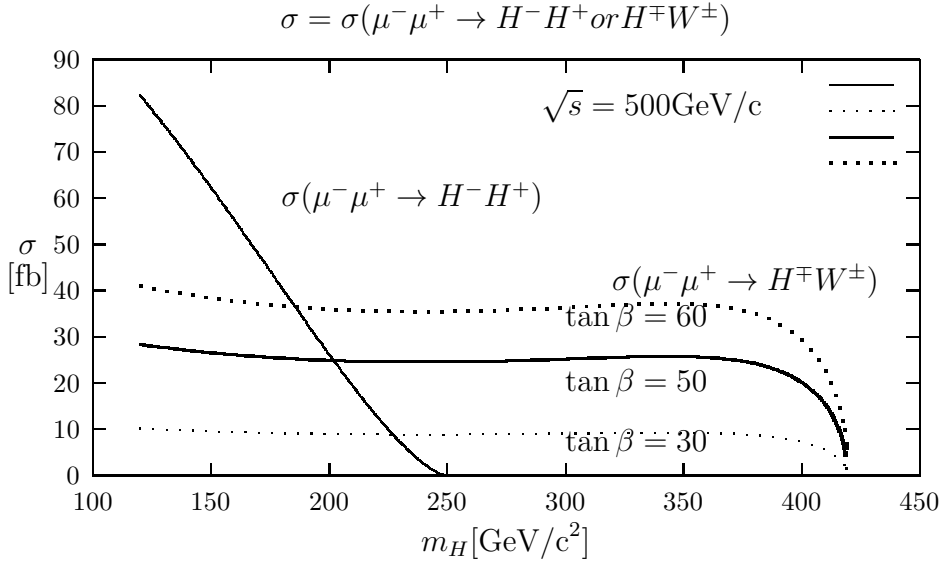}
\caption{Total cross section for the process
$\mu^- \mu^+ \rightarrow H^- H^+$ as a function
of $m_H$ compared with $\mu^- \mu^+ \rightarrow
H^{\mp} W^{\pm}$. We have taken
$\sqrt{s} = 500 \textrm{GeV/c}$ and $ \tan\beta= 30, 50, 60$.}
\label{sigmamumuhwhh_fig}
\end{center}
\end{figure}

In Figure \ref{sigmamumuhwhh_fig} we have plotted the total
cross section corresponding to the process $\mu^- \mu^+ \rightarrow
H^- H^+$ as a function of $m_H$ compared with $\mu^- \mu^+ \rightarrow
H^{\mp} W^{\pm}$. We have taken $\sqrt{s} = 500 \textrm{GeV/c}$.

\section{$\mu^- \mu^+ \rightarrow t \bar{t}$
annihilation}

The main background in the processes 
$\mu^- \mu^+ \rightarrow H^{\pm} W^{\mp}$
, assuming $H^+ \rightarrow t \bar{b}$
or $H^- \rightarrow \bar{t} b$ decays, comes from 
$t \bar{t}$ production.

\begin{figure}
\begin{center}
%\vspace*{-4.5cm}
%\scalebox{0.5}
{\includegraphics{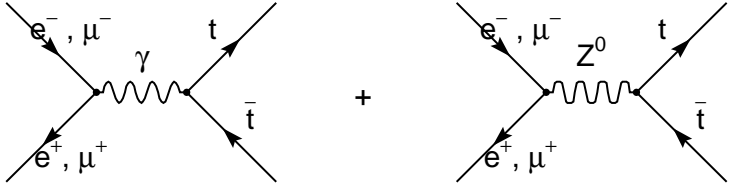}}
%\vspace*{0.7cm} 
\caption{Feynman diagrams corresponding to 
$  e^- e^+
\rightarrow t \bar{t}$ and $ \mu^- \mu^+
\rightarrow t \bar{t}$ annihilation.}
\label{mumuttbar}
\end{center}
\end{figure}

To lowest order in $e^2$ the Feynman diagrams corresponding
to the process $\mu^- \mu^+ \rightarrow t \bar{t}$  
are given in Figure \ref{mumuttbar}. The corresponding total cross
section is (see reference \cite{CM}):

\begin{eqnarray}
\lefteqn{
\frac{\sigma \left( \mu^- \mu^+ \rightarrow t \bar{t} \right)}
{\sigma_0} = \frac{3}{4} \{ m_t^2 s \{
\left[\frac{4}{3s} + \frac{\left( \frac{8}{3} \sin^2\theta_W
-1 \right)}{2\cos^2\theta_W \left( s - m_Z^2 \right)} \right]^2
}
\nonumber \\ & &
+ \left[ \frac{4}{3s} - \frac{\left(\frac{14}{3} \sin^2\theta_W
- \frac{16}{3} \sin^4\theta_W -1 \right)}{\sin^2\left
(2\theta_W\right) \left(s - m_Z^2 \right)} \right]^2 \}
\nonumber \\ & &
+ 2\left[ \frac{2}{3} + \frac{\left( \frac{8}{3}
\sin^2\theta_W -1 \right)s}{4 \cos^2\theta_W
\left( s - m_Z^2 \right)}\right]^2
+ \left[ \frac{s^2\left( 1 - \frac{4 m_t^2}{s} \right)}
{8 \cos^4\theta_W\left( s - m_Z^2\right)^2}\right]
\nonumber \\ & &
+ 2\left[ \frac{2}{3} + \frac{\left(2\sin^2\theta_W -1 \right)
\left(\frac{4}{3}\sin^2\theta_W - \frac{1}{2}\right)s}
{\sin^2\left(2\theta_W\right) \left(s - m_Z^2\right)}\right]^2
\nonumber \\ & &
+ \frac{\left(1 - \frac{4 m_t^2}{s} \right)
\left(2 \sin^2\theta_W -1 \right)^2 s^2}
{2\sin^4\left(2\theta_W\right) \left(s - m_Z^2\right)^2}
\} \left( 1 - \frac{4 m_t^2}{s} \right)^{1/2}
\label{mumuttbarsigma}
\end{eqnarray}

\noindent where

\begin{equation}
\sigma_0 = \sigma\left( e^- e^+ \rightarrow \mu^- \mu^+ 
\right) = \frac{4 \pi \alpha_{em}^2}{3s}
\end{equation}

In (\ref{mumuttbarsigma}) we have neglected $\Gamma_Z$ that is very small
for large values of $\sqrt{s}$. 

Taking $\sin^2\theta_W = 0.231$, $m_Z = 91.1876 \textrm{GeV/c}^2$,
$m_t= 174.3 \textrm{GeV/c}^2$ and
$\sqrt{s} = 500 \textrm{GeV/c}$ we get $\sigma\left(\mu^- \mu^+
\rightarrow t \bar{t}\right) = 495.1 \textrm{fb}$.

The total cross section corresponding to $e^- e^+ \rightarrow t \bar{t}$ 
is given by the same Equation (\ref{mumuttbarsigma}).

\section{$H^{\mp}W^{\pm}$ production at a Hadron Collider}
\subsection{$q \bar{q} \rightarrow H^- W^+ $ interaction}
From the Feynman diagrams of Figures \ref{qqHW_Feynman}
and \ref{qqHW_Feynman2} we obtain:

\begin{eqnarray}
\lefteqn{
\frac{d\sigma_I}{d\hat{t}}\left( q \bar{q}
\rightarrow H^- W^+ \right) = \frac{G_F^2}{48 \pi
\hat{s}} \{ m_q^2 \Lambda \left( \hat{s}, m_H^2, m_W^2 \right)
\left[ (\hat{C_{Hb}})^2 + (\hat{C_{Ab}})^2 \right]
}
\nonumber \\ & &
+2 \sum_{i,j = u,c,t} V_{iq}V_{jq}^*
[ m_q^2 c_{t_{1i}}c_{t_{1j}} \left( \hat{t}^2 +
\frac{\Lambda \left( \hat{s}, m_H^2, m_W^2 \right)
\sin^2\theta m_W^2}{2 \hat{s}} \right)
\nonumber \\ & &
+ m_i^2 m_j^2 c_{t_{2i}}c_{t_{2j}} 
\left( 2 m_W^2 + \frac{\Lambda 
\left( \hat{s}, m_H^2, m_W^2 \right) \sin^2\theta}
{4 \hat{s}} \right) ] + m_q^2 [ - 2 m_H^2m_W^2 
+ 2 \hat{t}^2
\nonumber \\ & &
+ \frac{1}{2} \Lambda \left( \hat{s}, m_H^2, m_W^2 \right)
\sin^2\theta ] \left( \hat{C_{Hb}} + \hat{C_{Ab}} \right)
\sum_{i= u, c, t} \Re{(V_{iq})} c_{t_{1i}} \}
\label{diffqqbar_H-W+}
\end{eqnarray}

\noindent for $ q = d, s, b$. 
In Equation (\ref{diffqqbar_H-W+}),
{ $\hat{C_{Hb}}$ and $\hat{C_{Ab}}$ are given by Equations
(\ref{CHb}) and (\ref{CAb}) replacing $s$ by $\hat{s}$.
$V_{iq}$ are elements of the CKM matrix.

\begin{equation}
c_{t_{1i}} = \frac{\tan\beta}{\left(\hat{t} - m_i^2\right)},
\label{ct1i}
\end{equation}

\noindent and

\begin{equation}
c_{t_{2i}} = \frac{\cot\beta}{\left( \hat{t} 
- m_i^2 \right)}.
\label{ct2i}
\end{equation}

On the other hand,

\begin{eqnarray}
\lefteqn{
\frac{d\sigma_{II}}{d\hat{t}}\left( q \bar{q}
\rightarrow H^- W^+ \right) = \frac{G_F^2}{48 \pi
\hat{s}} \{ m_q^2 \Lambda \left( \hat{s}, m_H^2, m_W^2 \right)
\left( \hat{C_{Ht}}^2 + \hat{C_{At}}^2 \right)
}
\nonumber \\ & &
+2 \sum_{i,j = d,s,b} V_{qi}^*V_{qj}
[ m_q^2 c_{u_{2i}}c_{u_{2j}} \left( \hat{u}^2 +
\frac{\Lambda \left( \hat{s}, m_H^2, m_W^2 \right) 
\sin^2\theta m_W^2}{2 \hat{s}} \right)
\nonumber \\ & &
+ m_i^2 m_j^2 c_{u_{1i}}c_{u_{1j}}
\left( 2 m_W^2 + \frac{\Lambda
\left( \hat{s}, m_H^2, m_W^2 \right) \sin^2\theta}
{4 \hat{s}} \right) ] + m_q^2 [ - 2 m_H^2m_W^2
+ 2 \hat{u}^2
\nonumber \\ & &
+ \frac{1}{2} \Lambda \left( \hat{s}, m_H^2, m_W^2 \right)
\sin^2\theta ] \left( \hat{C_{At}} - \hat{C_{Ht}} \right)
\sum_{i= d, s, b} \Re{(V_{qi})} c_{u_{2i}} \}
\label{diffqqbarIII_H-W+}
\end{eqnarray}

\noindent for $q = u,c$.

\begin{eqnarray}
\lefteqn{
\hat{C_{Ht}} = [ \frac{\left( \frac{1}{2} \sin2\alpha
-\sin^2\alpha \left(\tan\beta \right)^{-1} \right)}
{\left( \hat{s} - m_{H^0}^2 \right)}
}
\nonumber \\ & &
- \frac{ \left(\frac{1}{2} \sin2\alpha + \cos^2\alpha
\left( \tan\beta \right)^{-1}\right)}
{\left( \hat{s} - m_{h^0}^2 \right)} ],
\label{CHt}
\end{eqnarray}

\begin{equation}
\hat{C_{At}} = \frac{\cot\beta}{\left( \hat{s} - m_{A^0}^2
\right) },
\label{CAt}
\end{equation}

\begin{equation}
C_{u_{1i}} = \frac{\tan\beta}{\left(\hat{u} - m_i^2 \right)}
\label{Cu1i}
\end{equation}

\noindent and

\begin{equation}
C_{u_{2i}} = \frac{\cot\beta}{\left(\hat{u} - m_i^2 \right)}.
\label{Cu2i}  
\end{equation}

\begin{figure}
\begin{center}
%\vspace*{-4.5cm}
%\scalebox{0.5}
{\includegraphics{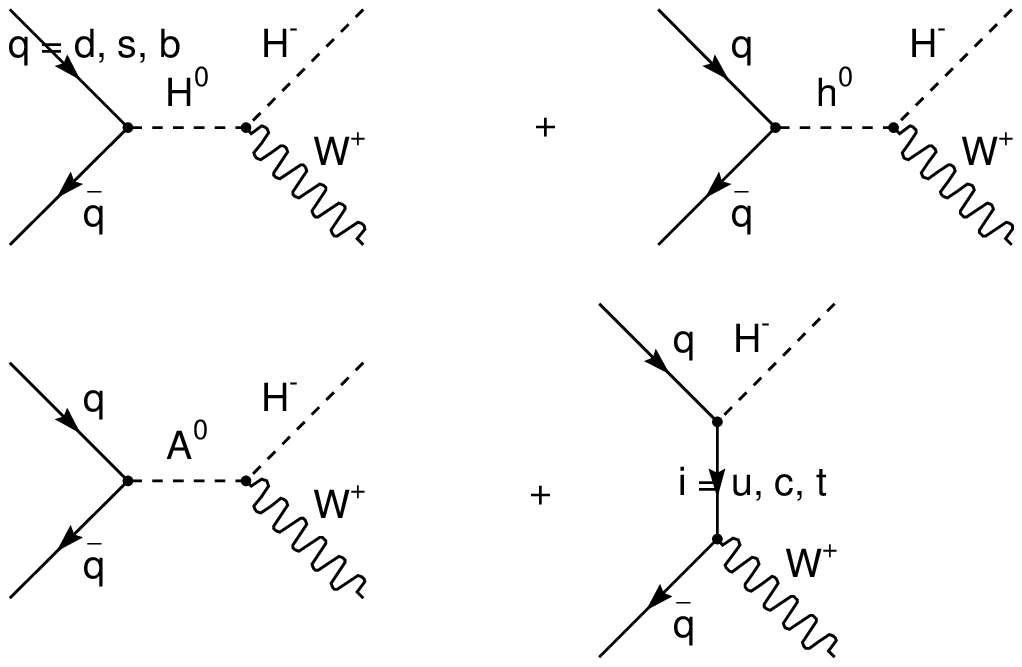}}
%\vspace*{0.7cm}
\caption{Feynman diagrams corresponding to the process
$\left( q \bar{q}
\rightarrow H^- W^+ \right)$ for $q = d, s, b$.}
\label{qqHW_Feynman}
\end{center}
\end{figure}

\begin{figure}
\begin{center}
%\vspace*{-4.5cm}
%\scalebox{0.5}
{\includegraphics{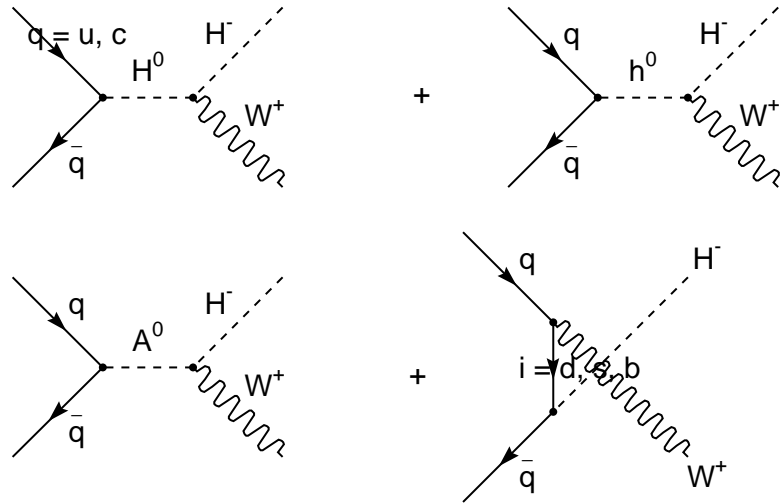}}
%\vspace*{0.7cm}
\caption{Feynman diagrams corresponding to the process
$\left( q \bar{q}
\rightarrow H^- W^+ \right)$ for $q = u, c$.}
\label{qqHW_Feynman2}
\end{center}
\end{figure}

The differential cross section corresponding to the process
$q \bar{q} \rightarrow H^+ W^-$ for $q = d, s, b$ is obtained
from Equation (\ref{diffqqbar_H-W+}) with the replacement $\hat{t}
\rightarrow \hat{u}$. For $ q = u, c$ we change $\hat{u}$ by
$\hat{t}$ in (\ref{diffqqbarIII_H-W+}).

\subsection{ $g g \rightarrow H^- W^+$ interaction}
The differential cross section corresponding to the sum of the
triangle diagrams in Figure \ref{gghw_Feynman} is given by:

\begin{eqnarray}
\lefteqn{
\frac{d \sigma_{\triangle}}{d\hat{t}} \left( g g \rightarrow
H^- W^+ \right) = \frac{\alpha_s^2 G_F^2}
{4096 \pi^3} 
}
\nonumber \\ & &
\times \Lambda \left( \hat{s}, m_H^2, m_W^2 \right)
\{ \left| \sum_{i = b,t} \left[\hat{C_{Hi}}\left(2\tau_i +
\tau_i \left( \tau_i - 1 \right) f(\tau_i)\right) \right]
\right|^2 
\nonumber \\ & &
+ \frac{1}{2} \left|\sum_{i = b,t} \hat{C_{Ai}} \tau_i 
f(\tau_i) \right|^2 \}
\label{ggH-W+_triangle}
\end{eqnarray} 

\noindent where 
\begin{equation}
\tau_i = \frac{4 m_i^2}{\hat{s}}
\end{equation}

\noindent and

\begin{equation}
f(\tau_i) = \left\{ \begin{array}{ll}
-2 \left[ \arcsin \left( \tau_i^{-1/2} \right) \right]^2 &
\mbox{if $\tau_i > 1$} \\
\frac{1}{2} \left[ \ln \left( \frac{1 + \left( 1- \tau_i \right)^{1/2}}
{1 - \left( 1 - \tau_i \right)^{1/2}} \right) - i \pi \right]^2 &
\mbox{if $\tau_i \le 1$}
\end{array} \right.
\label{ftau}
\end{equation}

\begin{figure}
\begin{center}
%\vspace*{-4.5cm}
%\scalebox{0.5}
{\includegraphics{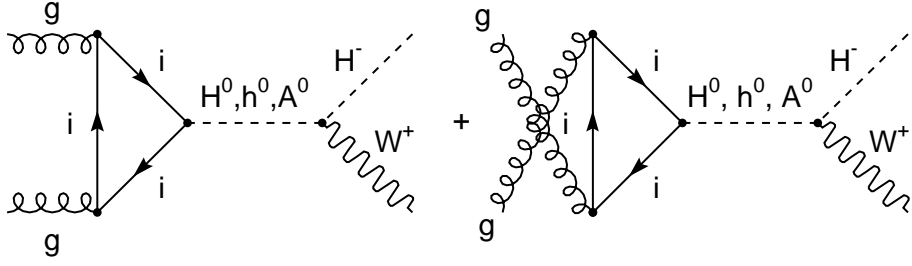}}
%\vspace*{0.7cm}
\caption{Triangle diagrams corresponding to the process
$ g g \rightarrow H^- W^+ $. $i = b, t$.}
\label{gghw_Feynman}
\end{center}
\end{figure}

Due to charge-conjugation invariance 

\begin{equation}
\frac{d \sigma_{\triangle}}{d\hat{t}} \left( g g \rightarrow
H^- W^+ \right) = \frac{d \sigma_{\triangle}}{d\hat{t}} \left( g g 
\rightarrow
H^+ W^- \right). 
\end{equation}

Equations (\ref{diffqqbar_H-W+}), (\ref{diffqqbarIII_H-W+})
and (\ref{ggH-W+_triangle}) are in agreement with the differential cross
sections calculated in reference \cite{Barrientos_K}. 
In this reference, the differential cross section 
corresponding to the sum of the box diagrams of Figures
\ref{gghwboxI} and \ref{gghwboxII}, also has been calculated 
with the aid of the computer packages FEYNARTS, FEYNCALC and FF.
According to the analisis presented in \cite{Barrientos_K},
the dominant subprocesses of $W^{\pm}H^{\mp}$ associated production
are $b\bar{b} \rightarrow W^{\pm}H^{\mp}$ at the tree level and 
$g g \rightarrow W^{\pm} H^{\mp}$ at one loop.

\begin{figure}
\begin{center}
%\vspace*{-4.5cm}
%\scalebox{0.5}
{\includegraphics{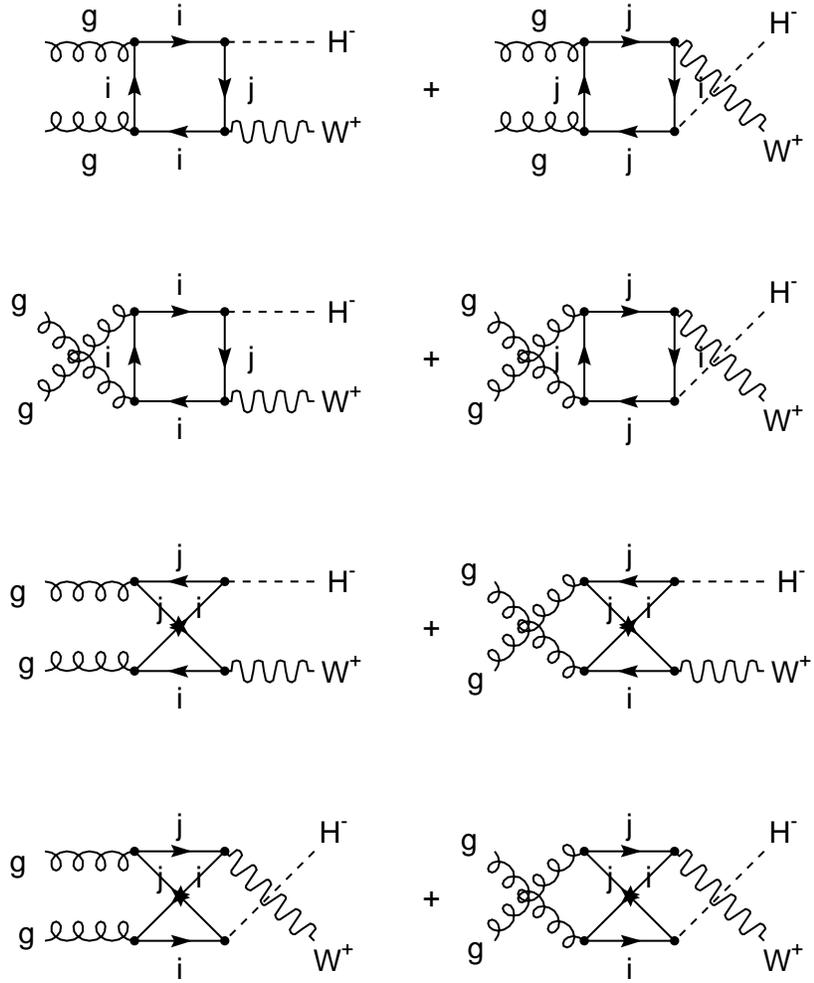}}
%\vspace*{0.7cm}
\caption{Box diagrams corresponding to the process
$ g g \rightarrow H^- W^+ $. $i = d,s,b$ ; $j=u,c,t$.
Continued in Figure \ref{gghwboxII}.}
\label{gghwboxI}
\end{center}  
\end{figure}

\begin{figure}
\begin{center}
%\vspace*{-4.5cm}
%\scalebox{0.5}
{\includegraphics{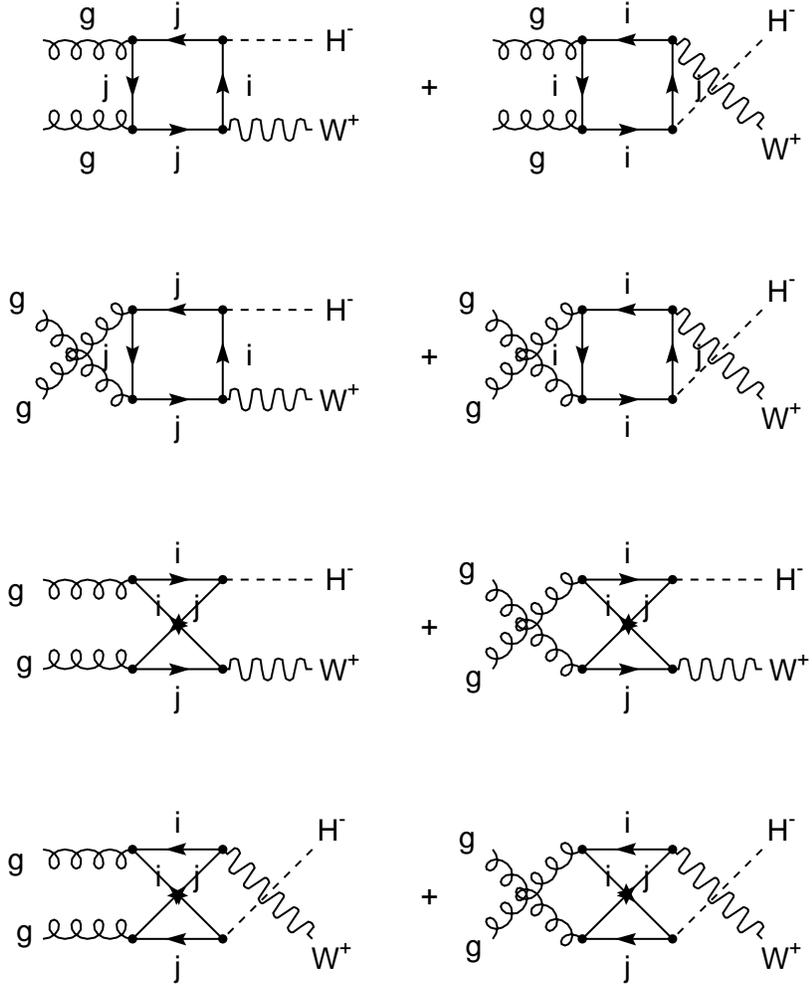}}
%\vspace*{0.7cm}
\caption{
Continued from Figure \ref{gghwboxI}.}
\label{gghwboxII}
\end{center}
\end{figure}
\subsection{Differential cross section $p \bar{p}
\rightarrow H^{\mp} W^{\pm} X$}

The differential cross section corresponding to the channel
$p \bar{p} \rightarrow H^{\mp} W^{\pm} X$ is:

\begin{eqnarray}
\frac{d^2 \sigma}{dy d \left( p_T \right)^2}
\left(p \bar{p}
\rightarrow H^{\mp} W^{\pm} X\right) & = & 
\sum_{f} { \int_{x_{amin}}^1 {
dx_a f_f \left( x_a, m_a^2 \right) f_f \left( x_b, m_b^2 \right)
}}
\nonumber \\ & &
\times\frac{x_b \hat{s}}{\left(m_H^2 - \hat{u}\right)}
\frac{d \sigma}{d \hat{t}} \left( f \bar{f} \rightarrow 
H^{\mp} W^{\pm} 
\right)
\label{pp_HWX}  
\end{eqnarray}
where $f$ is $q$ or $g$,
\begin{equation}
x_{amin} = \frac{\sqrt{s} m_T e^y + m_H^2 - m_W^2}
{s - \sqrt{s} m_T e^{-y}},
\label{xaminHW}
\end{equation}
\begin{equation}
m_T = \left( m_W^2 + p_T^2 \right)^{\frac{1}{2}},
\label{mTW}
\end{equation}
\begin{equation}
x_b = \frac{x_a \sqrt{s} m_T e^{-y} + m_H^2 - m_W^2}
{x_a s - \sqrt{s} m_T e^y},
\label{xbHW}
\end{equation}
\begin{equation}
\hat{s} = x_a x_b s,
\label{s_hatHW}
\end{equation}
\begin{equation}
p_T^2 = \frac{\Lambda \left( \hat{s}, m_H^2, m_W^2 \right) \sin^2 
\theta}
{4 \hat{s}},
\label{pT2HW}
\end{equation}
\begin{equation}
\hat{u} = \frac{1}{2} \left( m_H^2 + m_W^2 -\hat{s}
- \cos \theta \Lambda^{1/2} ( \hat{s}, m_H^2, m_W^2 )
\right),
\label{uhatHW}  
\end{equation}
\begin{equation}
\hat{t} = \frac{1}{2} \left( m_H^2 + m_W^2 -\hat{s}
+ \cos \theta \Lambda^{1/2} ( \hat{s}, m_H^2, m_W^2 )
\right),
\label{thatHW}  
\end{equation}
\begin{equation}
\hat{u} \hat{t} = m_H^2 m_W^2 + \hat{s} p_T^2,
\label{utHW}
\end{equation}  
and
\begin{equation}
\cos\theta = \left( 1 - \frac{4 \hat{s} p_T^2}
{ \Lambda^{1/2} ( \hat{s}, m_H^2, m_W^2 )}
\right)^{1/2}
\label{costhetaHW}
\end{equation}

\noindent $y$ is the rapidity of $W^{\pm}$, $\theta$ is the angle of
dispersion in the center of mass system,
$p_T$ is the transverse 
momentum of $W^{\pm}$, $f_f$ are the unpolarized parton 
distribution functions for quarks (antiquarks) or gluons.
Finally, $m_a^2$ or $m_b^2$ represent the factorization scale.

A similar expression is valid for the reaction $p p \rightarrow
H^{\mp} W^{\pm} + X$.

In Figure \ref{LHC_figure} (taken from reference \cite{Barrientos_K})
the total cross section $\sigma$ of $pp \rightarrow W^{\pm} H^{\mp} + X$
via $b\bar{b}$ annihilation and $gg$ fusion is plotted as a function of
$m_H$ at LHC energies ($\sqrt{s} = 14 \textrm{TeV/c}$) for $\tan\beta 
=30$. Other 
contributions are negligible.

In Figure \ref{Tevatron_figure} (taken from reference 
\cite{Barrientos_K})
the total cross section $\sigma$ of $p\bar{p} \rightarrow W^{\pm} H^{\mp} 
+ X$
via $b\bar{b}$ annihilation and $gg$ fusion is plotted as a function of
$m_H$ at the Tevatron energy ($\sqrt{s} = 2 \textrm{TeV/c}$) for 
$\tan\beta 
=30$. The
contributions of the other partons are negligible.
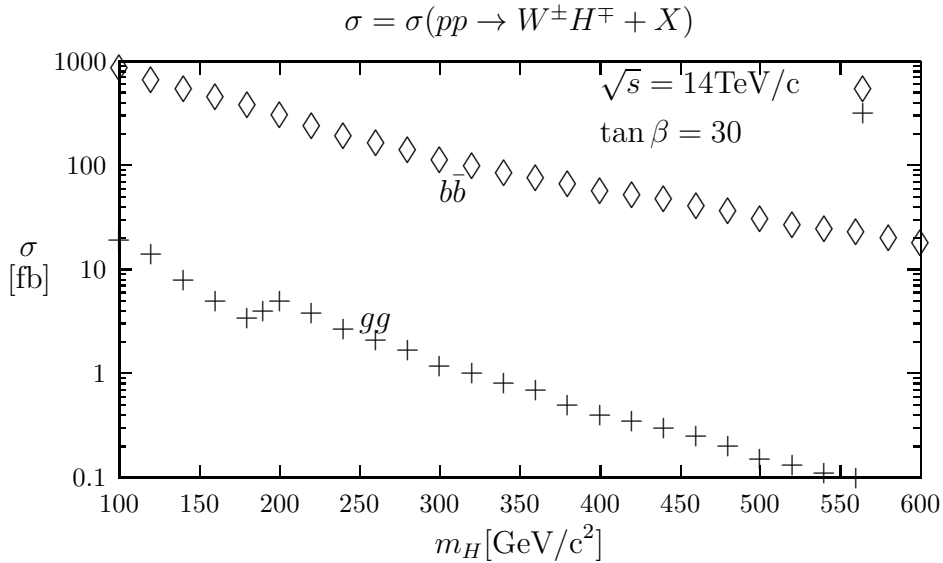
\begin{figure}
\begin{center}
% GNUPLOT: LaTeX picture
\setlength{\unitlength}{0.240900pt}
\ifx\plotpoint\undefined\newsavebox{\plotpoint}\fi
\sbox{\plotpoint}{\rule[-0.200pt]{0.400pt}{0.400pt}}%
\begin{picture}(1500,900)(0,0)
\font\gnuplot=cmr10 at 10pt
\gnuplot
\sbox{\plotpoint}{\rule[-0.200pt]{0.400pt}{0.400pt}}%
\put(181.0,123.0){\rule[-0.200pt]{4.818pt}{0.400pt}}
\put(161,123){\makebox(0,0)[r]{0.1}}
\put(1419.0,123.0){\rule[-0.200pt]{4.818pt}{0.400pt}}
\put(181.0,172.0){\rule[-0.200pt]{2.409pt}{0.400pt}}
\put(1429.0,172.0){\rule[-0.200pt]{2.409pt}{0.400pt}}
\put(181.0,201.0){\rule[-0.200pt]{2.409pt}{0.400pt}}
\put(1429.0,201.0){\rule[-0.200pt]{2.409pt}{0.400pt}}
\put(181.0,221.0){\rule[-0.200pt]{2.409pt}{0.400pt}}
\put(1429.0,221.0){\rule[-0.200pt]{2.409pt}{0.400pt}}
\put(181.0,237.0){\rule[-0.200pt]{2.409pt}{0.400pt}}
\put(1429.0,237.0){\rule[-0.200pt]{2.409pt}{0.400pt}}
\put(181.0,250.0){\rule[-0.200pt]{2.409pt}{0.400pt}}
\put(1429.0,250.0){\rule[-0.200pt]{2.409pt}{0.400pt}}
\put(181.0,261.0){\rule[-0.200pt]{2.409pt}{0.400pt}}
\put(1429.0,261.0){\rule[-0.200pt]{2.409pt}{0.400pt}}
\put(181.0,271.0){\rule[-0.200pt]{2.409pt}{0.400pt}}
\put(1429.0,271.0){\rule[-0.200pt]{2.409pt}{0.400pt}}
\put(181.0,279.0){\rule[-0.200pt]{2.409pt}{0.400pt}}
\put(1429.0,279.0){\rule[-0.200pt]{2.409pt}{0.400pt}}
\put(181.0,287.0){\rule[-0.200pt]{4.818pt}{0.400pt}}
\put(161,287){\makebox(0,0)[r]{1}}
\put(1419.0,287.0){\rule[-0.200pt]{4.818pt}{0.400pt}}
\put(181.0,336.0){\rule[-0.200pt]{2.409pt}{0.400pt}}
\put(1429.0,336.0){\rule[-0.200pt]{2.409pt}{0.400pt}}
\put(181.0,365.0){\rule[-0.200pt]{2.409pt}{0.400pt}}
\put(1429.0,365.0){\rule[-0.200pt]{2.409pt}{0.400pt}}
\put(181.0,385.0){\rule[-0.200pt]{2.409pt}{0.400pt}}
\put(1429.0,385.0){\rule[-0.200pt]{2.409pt}{0.400pt}}
\put(181.0,401.0){\rule[-0.200pt]{2.409pt}{0.400pt}}
\put(1429.0,401.0){\rule[-0.200pt]{2.409pt}{0.400pt}}
\put(181.0,414.0){\rule[-0.200pt]{2.409pt}{0.400pt}}
\put(1429.0,414.0){\rule[-0.200pt]{2.409pt}{0.400pt}}
\put(181.0,425.0){\rule[-0.200pt]{2.409pt}{0.400pt}}
\put(1429.0,425.0){\rule[-0.200pt]{2.409pt}{0.400pt}}
\put(181.0,434.0){\rule[-0.200pt]{2.409pt}{0.400pt}}
\put(1429.0,434.0){\rule[-0.200pt]{2.409pt}{0.400pt}}
\put(181.0,443.0){\rule[-0.200pt]{2.409pt}{0.400pt}}
\put(1429.0,443.0){\rule[-0.200pt]{2.409pt}{0.400pt}}
\put(181.0,450.0){\rule[-0.200pt]{4.818pt}{0.400pt}}
\put(161,450){\makebox(0,0)[r]{10}}
\put(1419.0,450.0){\rule[-0.200pt]{4.818pt}{0.400pt}}
\put(181.0,499.0){\rule[-0.200pt]{2.409pt}{0.400pt}}
\put(1429.0,499.0){\rule[-0.200pt]{2.409pt}{0.400pt}}
\put(181.0,528.0){\rule[-0.200pt]{2.409pt}{0.400pt}}
\put(1429.0,528.0){\rule[-0.200pt]{2.409pt}{0.400pt}}
\put(181.0,548.0){\rule[-0.200pt]{2.409pt}{0.400pt}}
\put(1429.0,548.0){\rule[-0.200pt]{2.409pt}{0.400pt}}
\put(181.0,564.0){\rule[-0.200pt]{2.409pt}{0.400pt}}
\put(1429.0,564.0){\rule[-0.200pt]{2.409pt}{0.400pt}}
\put(181.0,577.0){\rule[-0.200pt]{2.409pt}{0.400pt}}
\put(1429.0,577.0){\rule[-0.200pt]{2.409pt}{0.400pt}}
\put(181.0,588.0){\rule[-0.200pt]{2.409pt}{0.400pt}}
\put(1429.0,588.0){\rule[-0.200pt]{2.409pt}{0.400pt}}
\put(181.0,598.0){\rule[-0.200pt]{2.409pt}{0.400pt}}
\put(1429.0,598.0){\rule[-0.200pt]{2.409pt}{0.400pt}}
\put(181.0,606.0){\rule[-0.200pt]{2.409pt}{0.400pt}}
\put(1429.0,606.0){\rule[-0.200pt]{2.409pt}{0.400pt}}
\put(181.0,614.0){\rule[-0.200pt]{4.818pt}{0.400pt}}
\put(161,614){\makebox(0,0)[r]{100}}
\put(1419.0,614.0){\rule[-0.200pt]{4.818pt}{0.400pt}}
\put(181.0,663.0){\rule[-0.200pt]{2.409pt}{0.400pt}}
\put(1429.0,663.0){\rule[-0.200pt]{2.409pt}{0.400pt}}
\put(181.0,692.0){\rule[-0.200pt]{2.409pt}{0.400pt}}
\put(1429.0,692.0){\rule[-0.200pt]{2.409pt}{0.400pt}}
\put(181.0,712.0){\rule[-0.200pt]{2.409pt}{0.400pt}}
\put(1429.0,712.0){\rule[-0.200pt]{2.409pt}{0.400pt}}
\put(181.0,728.0){\rule[-0.200pt]{2.409pt}{0.400pt}}
\put(1429.0,728.0){\rule[-0.200pt]{2.409pt}{0.400pt}}
\put(181.0,741.0){\rule[-0.200pt]{2.409pt}{0.400pt}}
\put(1429.0,741.0){\rule[-0.200pt]{2.409pt}{0.400pt}}
\put(181.0,752.0){\rule[-0.200pt]{2.409pt}{0.400pt}}
\put(1429.0,752.0){\rule[-0.200pt]{2.409pt}{0.400pt}}
\put(181.0,761.0){\rule[-0.200pt]{2.409pt}{0.400pt}}
\put(1429.0,761.0){\rule[-0.200pt]{2.409pt}{0.400pt}}
\put(181.0,770.0){\rule[-0.200pt]{2.409pt}{0.400pt}}
\put(1429.0,770.0){\rule[-0.200pt]{2.409pt}{0.400pt}}
\put(181.0,777.0){\rule[-0.200pt]{4.818pt}{0.400pt}}
\put(161,777){\makebox(0,0)[r]{1000}}
\put(1419.0,777.0){\rule[-0.200pt]{4.818pt}{0.400pt}}
\put(181.0,123.0){\rule[-0.200pt]{0.400pt}{4.818pt}}
\put(181,82){\makebox(0,0){100}}
\put(181.0,757.0){\rule[-0.200pt]{0.400pt}{4.818pt}}
\put(307.0,123.0){\rule[-0.200pt]{0.400pt}{4.818pt}}
\put(307,82){\makebox(0,0){150}}
\put(307.0,757.0){\rule[-0.200pt]{0.400pt}{4.818pt}}
\put(433.0,123.0){\rule[-0.200pt]{0.400pt}{4.818pt}}
\put(433,82){\makebox(0,0){200}}
\put(433.0,757.0){\rule[-0.200pt]{0.400pt}{4.818pt}}
\put(558.0,123.0){\rule[-0.200pt]{0.400pt}{4.818pt}}
\put(558,82){\makebox(0,0){250}}
\put(558.0,757.0){\rule[-0.200pt]{0.400pt}{4.818pt}}
\put(684.0,123.0){\rule[-0.200pt]{0.400pt}{4.818pt}}
\put(684,82){\makebox(0,0){300}}
\put(684.0,757.0){\rule[-0.200pt]{0.400pt}{4.818pt}}
\put(810.0,123.0){\rule[-0.200pt]{0.400pt}{4.818pt}}
\put(810,82){\makebox(0,0){350}}
\put(810.0,757.0){\rule[-0.200pt]{0.400pt}{4.818pt}}
\put(936.0,123.0){\rule[-0.200pt]{0.400pt}{4.818pt}}
\put(936,82){\makebox(0,0){400}}
\put(936.0,757.0){\rule[-0.200pt]{0.400pt}{4.818pt}}
\put(1062.0,123.0){\rule[-0.200pt]{0.400pt}{4.818pt}}
\put(1062,82){\makebox(0,0){450}}
\put(1062.0,757.0){\rule[-0.200pt]{0.400pt}{4.818pt}}
\put(1187.0,123.0){\rule[-0.200pt]{0.400pt}{4.818pt}}
\put(1187,82){\makebox(0,0){500}}
\put(1187.0,757.0){\rule[-0.200pt]{0.400pt}{4.818pt}}
\put(1313.0,123.0){\rule[-0.200pt]{0.400pt}{4.818pt}}
\put(1313,82){\makebox(0,0){550}}
\put(1313.0,757.0){\rule[-0.200pt]{0.400pt}{4.818pt}}
\put(1439.0,123.0){\rule[-0.200pt]{0.400pt}{4.818pt}}
\put(1439,82){\makebox(0,0){600}}
\put(1439.0,757.0){\rule[-0.200pt]{0.400pt}{4.818pt}}
\put(181.0,123.0){\rule[-0.200pt]{303.052pt}{0.400pt}}
\put(1439.0,123.0){\rule[-0.200pt]{0.400pt}{157.549pt}}
\put(181.0,777.0){\rule[-0.200pt]{303.052pt}{0.400pt}}
\put(40,450){\makebox(0,0){\shortstack{$\sigma$ \\ $[\textrm{fb}]$}}}
\put(810,21){\makebox(0,0){$m_H [\textrm{GeV/c}^2]$}}
\put(810,839){\makebox(0,0){$\sigma = \sigma(pp \rightarrow W^{\pm} H^{\mp} + X)$}}
\put(936,741){\makebox(0,0)[l]{$\sqrt{s} = 14 \textrm{TeV/c}$}}
\put(936,663){\makebox(0,0)[l]{$\tan\beta=30$}}
\put(684,577){\makebox(0,0)[l]{$b\bar{b}$}}
\put(558,365){\makebox(0,0)[l]{$gg$}}
\put(181.0,123.0){\rule[-0.200pt]{0.400pt}{157.549pt}}
\put(1279,737){\makebox(0,0)[r]{ }}
\put(181,770){\raisebox{-.8pt}{\makebox(0,0){$\Diamond$}}}
\put(231,752){\raisebox{-.8pt}{\makebox(0,0){$\Diamond$}}}
\put(282,738){\raisebox{-.8pt}{\makebox(0,0){$\Diamond$}}}
\put(332,725){\raisebox{-.8pt}{\makebox(0,0){$\Diamond$}}}
\put(382,712){\raisebox{-.8pt}{\makebox(0,0){$\Diamond$}}}
\put(433,696){\raisebox{-.8pt}{\makebox(0,0){$\Diamond$}}}
\put(483,679){\raisebox{-.8pt}{\makebox(0,0){$\Diamond$}}}
\put(533,663){\raisebox{-.8pt}{\makebox(0,0){$\Diamond$}}}
\put(584,653){\raisebox{-.8pt}{\makebox(0,0){$\Diamond$}}}
\put(634,642){\raisebox{-.8pt}{\makebox(0,0){$\Diamond$}}}
\put(684,626){\raisebox{-.8pt}{\makebox(0,0){$\Diamond$}}}
\put(735,617){\raisebox{-.8pt}{\makebox(0,0){$\Diamond$}}}
\put(785,606){\raisebox{-.8pt}{\makebox(0,0){$\Diamond$}}}
\put(835,598){\raisebox{-.8pt}{\makebox(0,0){$\Diamond$}}}
\put(885,588){\raisebox{-.8pt}{\makebox(0,0){$\Diamond$}}}
\put(936,577){\raisebox{-.8pt}{\makebox(0,0){$\Diamond$}}}
\put(986,571){\raisebox{-.8pt}{\makebox(0,0){$\Diamond$}}}
\put(1036,564){\raisebox{-.8pt}{\makebox(0,0){$\Diamond$}}}
\put(1087,554){\raisebox{-.8pt}{\makebox(0,0){$\Diamond$}}}
\put(1137,545){\raisebox{-.8pt}{\makebox(0,0){$\Diamond$}}}
\put(1187,533){\raisebox{-.8pt}{\makebox(0,0){$\Diamond$}}}
\put(1238,523){\raisebox{-.8pt}{\makebox(0,0){$\Diamond$}}}
\put(1288,518){\raisebox{-.8pt}{\makebox(0,0){$\Diamond$}}}
\put(1338,512){\raisebox{-.8pt}{\makebox(0,0){$\Diamond$}}}
\put(1389,503){\raisebox{-.8pt}{\makebox(0,0){$\Diamond$}}}
\put(1439,496){\raisebox{-.8pt}{\makebox(0,0){$\Diamond$}}}
\put(1349,737){\raisebox{-.8pt}{\makebox(0,0){$\Diamond$}}}
\put(1279,696){\makebox(0,0)[r]{ }}
\put(181,496){\makebox(0,0){$+$}}
\put(231,474){\makebox(0,0){$+$}}
\put(282,434){\makebox(0,0){$+$}}
\put(332,401){\makebox(0,0){$+$}}
\put(382,373){\makebox(0,0){$+$}}
\put(407,385){\makebox(0,0){$+$}}
\put(433,401){\makebox(0,0){$+$}}
\put(483,381){\makebox(0,0){$+$}}
\put(533,357){\makebox(0,0){$+$}}
\put(584,339){\makebox(0,0){$+$}}
\put(634,324){\makebox(0,0){$+$}}
\put(684,299){\makebox(0,0){$+$}}
\put(735,287){\makebox(0,0){$+$}}
\put(785,271){\makebox(0,0){$+$}}
\put(835,261){\makebox(0,0){$+$}}
\put(885,237){\makebox(0,0){$+$}}
\put(936,221){\makebox(0,0){$+$}}
\put(986,212){\makebox(0,0){$+$}}
\put(1036,201){\makebox(0,0){$+$}}
\put(1087,188){\makebox(0,0){$+$}}
\put(1137,172){\makebox(0,0){$+$}}
\put(1187,152){\makebox(0,0){$+$}}
\put(1238,142){\makebox(0,0){$+$}}
\put(1288,130){\makebox(0,0){$+$}}
\put(1338,123){\makebox(0,0){$+$}}
\put(1349,696){\makebox(0,0){$+$}}
\end{picture}
\caption{Total cross section for the process
$p p \rightarrow W^{\pm} H^{\mp} + X$ as a function
of $m_H$ via $b\bar{b}$ annihilation and $gg$ fusion
at LHC energies 
($\sqrt{s} = 14 \textrm{TeV/c}$) for
$\tan\beta = 30$. Taken from \cite{Barrientos_K}.}
\label{LHC_figure}
\end{center}
\end{figure}
 
\begin{figure}
\begin{center}
% GNUPLOT: LaTeX picture
\setlength{\unitlength}{0.240900pt}
\ifx\plotpoint\undefined\newsavebox{\plotpoint}\fi
\sbox{\plotpoint}{\rule[-0.200pt]{0.400pt}{0.400pt}}%
\begin{picture}(1500,900)(0,0)
\font\gnuplot=cmr10 at 10pt
\gnuplot
\sbox{\plotpoint}{\rule[-0.200pt]{0.400pt}{0.400pt}}%
\put(201.0,123.0){\rule[-0.200pt]{4.818pt}{0.400pt}}
\put(181,123){\makebox(0,0)[r]{0.001}}
\put(1419.0,123.0){\rule[-0.200pt]{4.818pt}{0.400pt}}
\put(201.0,172.0){\rule[-0.200pt]{2.409pt}{0.400pt}}
\put(1429.0,172.0){\rule[-0.200pt]{2.409pt}{0.400pt}}
\put(201.0,201.0){\rule[-0.200pt]{2.409pt}{0.400pt}}
\put(1429.0,201.0){\rule[-0.200pt]{2.409pt}{0.400pt}}
\put(201.0,221.0){\rule[-0.200pt]{2.409pt}{0.400pt}}
\put(1429.0,221.0){\rule[-0.200pt]{2.409pt}{0.400pt}}
\put(201.0,237.0){\rule[-0.200pt]{2.409pt}{0.400pt}}
\put(1429.0,237.0){\rule[-0.200pt]{2.409pt}{0.400pt}}
\put(201.0,250.0){\rule[-0.200pt]{2.409pt}{0.400pt}}
\put(1429.0,250.0){\rule[-0.200pt]{2.409pt}{0.400pt}}
\put(201.0,261.0){\rule[-0.200pt]{2.409pt}{0.400pt}}
\put(1429.0,261.0){\rule[-0.200pt]{2.409pt}{0.400pt}}
\put(201.0,271.0){\rule[-0.200pt]{2.409pt}{0.400pt}}
\put(1429.0,271.0){\rule[-0.200pt]{2.409pt}{0.400pt}}
\put(201.0,279.0){\rule[-0.200pt]{2.409pt}{0.400pt}}
\put(1429.0,279.0){\rule[-0.200pt]{2.409pt}{0.400pt}}
\put(201.0,287.0){\rule[-0.200pt]{4.818pt}{0.400pt}}
\put(181,287){\makebox(0,0)[r]{0.01}}
\put(1419.0,287.0){\rule[-0.200pt]{4.818pt}{0.400pt}}
\put(201.0,336.0){\rule[-0.200pt]{2.409pt}{0.400pt}}
\put(1429.0,336.0){\rule[-0.200pt]{2.409pt}{0.400pt}}
\put(201.0,365.0){\rule[-0.200pt]{2.409pt}{0.400pt}}
\put(1429.0,365.0){\rule[-0.200pt]{2.409pt}{0.400pt}}
\put(201.0,385.0){\rule[-0.200pt]{2.409pt}{0.400pt}}
\put(1429.0,385.0){\rule[-0.200pt]{2.409pt}{0.400pt}}
\put(201.0,401.0){\rule[-0.200pt]{2.409pt}{0.400pt}}
\put(1429.0,401.0){\rule[-0.200pt]{2.409pt}{0.400pt}}
\put(201.0,414.0){\rule[-0.200pt]{2.409pt}{0.400pt}}
\put(1429.0,414.0){\rule[-0.200pt]{2.409pt}{0.400pt}}
\put(201.0,425.0){\rule[-0.200pt]{2.409pt}{0.400pt}}
\put(1429.0,425.0){\rule[-0.200pt]{2.409pt}{0.400pt}}
\put(201.0,434.0){\rule[-0.200pt]{2.409pt}{0.400pt}}
\put(1429.0,434.0){\rule[-0.200pt]{2.409pt}{0.400pt}}
\put(201.0,443.0){\rule[-0.200pt]{2.409pt}{0.400pt}}
\put(1429.0,443.0){\rule[-0.200pt]{2.409pt}{0.400pt}}
\put(201.0,450.0){\rule[-0.200pt]{4.818pt}{0.400pt}}
\put(181,450){\makebox(0,0)[r]{0.1}}
\put(1419.0,450.0){\rule[-0.200pt]{4.818pt}{0.400pt}}
\put(201.0,499.0){\rule[-0.200pt]{2.409pt}{0.400pt}}
\put(1429.0,499.0){\rule[-0.200pt]{2.409pt}{0.400pt}}
\put(201.0,528.0){\rule[-0.200pt]{2.409pt}{0.400pt}}
\put(1429.0,528.0){\rule[-0.200pt]{2.409pt}{0.400pt}}
\put(201.0,548.0){\rule[-0.200pt]{2.409pt}{0.400pt}}
\put(1429.0,548.0){\rule[-0.200pt]{2.409pt}{0.400pt}}
\put(201.0,564.0){\rule[-0.200pt]{2.409pt}{0.400pt}}
\put(1429.0,564.0){\rule[-0.200pt]{2.409pt}{0.400pt}}
\put(201.0,577.0){\rule[-0.200pt]{2.409pt}{0.400pt}}
\put(1429.0,577.0){\rule[-0.200pt]{2.409pt}{0.400pt}}
\put(201.0,588.0){\rule[-0.200pt]{2.409pt}{0.400pt}}
\put(1429.0,588.0){\rule[-0.200pt]{2.409pt}{0.400pt}}
\put(201.0,598.0){\rule[-0.200pt]{2.409pt}{0.400pt}}
\put(1429.0,598.0){\rule[-0.200pt]{2.409pt}{0.400pt}}
\put(201.0,606.0){\rule[-0.200pt]{2.409pt}{0.400pt}}
\put(1429.0,606.0){\rule[-0.200pt]{2.409pt}{0.400pt}}
\put(201.0,614.0){\rule[-0.200pt]{4.818pt}{0.400pt}}
\put(181,614){\makebox(0,0)[r]{1}}
\put(1419.0,614.0){\rule[-0.200pt]{4.818pt}{0.400pt}}
\put(201.0,663.0){\rule[-0.200pt]{2.409pt}{0.400pt}}
\put(1429.0,663.0){\rule[-0.200pt]{2.409pt}{0.400pt}}
\put(201.0,692.0){\rule[-0.200pt]{2.409pt}{0.400pt}}
\put(1429.0,692.0){\rule[-0.200pt]{2.409pt}{0.400pt}}
\put(201.0,712.0){\rule[-0.200pt]{2.409pt}{0.400pt}}
\put(1429.0,712.0){\rule[-0.200pt]{2.409pt}{0.400pt}}
\put(201.0,728.0){\rule[-0.200pt]{2.409pt}{0.400pt}}
\put(1429.0,728.0){\rule[-0.200pt]{2.409pt}{0.400pt}}
\put(201.0,741.0){\rule[-0.200pt]{2.409pt}{0.400pt}}
\put(1429.0,741.0){\rule[-0.200pt]{2.409pt}{0.400pt}}
\put(201.0,752.0){\rule[-0.200pt]{2.409pt}{0.400pt}}
\put(1429.0,752.0){\rule[-0.200pt]{2.409pt}{0.400pt}}
\put(201.0,761.0){\rule[-0.200pt]{2.409pt}{0.400pt}}
\put(1429.0,761.0){\rule[-0.200pt]{2.409pt}{0.400pt}}
\put(201.0,770.0){\rule[-0.200pt]{2.409pt}{0.400pt}}
\put(1429.0,770.0){\rule[-0.200pt]{2.409pt}{0.400pt}}
\put(201.0,777.0){\rule[-0.200pt]{4.818pt}{0.400pt}}
\put(181,777){\makebox(0,0)[r]{10}}
\put(1419.0,777.0){\rule[-0.200pt]{4.818pt}{0.400pt}}
\put(201.0,123.0){\rule[-0.200pt]{0.400pt}{4.818pt}}
\put(201,82){\makebox(0,0){100}}
\put(201.0,757.0){\rule[-0.200pt]{0.400pt}{4.818pt}}
\put(511.0,123.0){\rule[-0.200pt]{0.400pt}{4.818pt}}
\put(511,82){\makebox(0,0){150}}
\put(511.0,757.0){\rule[-0.200pt]{0.400pt}{4.818pt}}
\put(820.0,123.0){\rule[-0.200pt]{0.400pt}{4.818pt}}
\put(820,82){\makebox(0,0){200}}
\put(820.0,757.0){\rule[-0.200pt]{0.400pt}{4.818pt}}
\put(1130.0,123.0){\rule[-0.200pt]{0.400pt}{4.818pt}}
\put(1130,82){\makebox(0,0){250}}
\put(1130.0,757.0){\rule[-0.200pt]{0.400pt}{4.818pt}}
\put(1439.0,123.0){\rule[-0.200pt]{0.400pt}{4.818pt}}
\put(1439,82){\makebox(0,0){300}}
\put(1439.0,757.0){\rule[-0.200pt]{0.400pt}{4.818pt}}
\put(201.0,123.0){\rule[-0.200pt]{298.234pt}{0.400pt}}
\put(1439.0,123.0){\rule[-0.200pt]{0.400pt}{157.549pt}}
\put(201.0,777.0){\rule[-0.200pt]{298.234pt}{0.400pt}}
\put(40,450){\makebox(0,0){\shortstack{$\sigma$ \\ $[\textrm{fb}]$}}}
\put(820,21){\makebox(0,0){$m_H [\textrm{GeV/c}^2]$}}
\put(820,839){\makebox(0,0){$\sigma = \sigma(p \bar{p} \rightarrow W^{\pm} H^{\mp} + X)$}}
\put(1068,728){\makebox(0,0)[l]{$\sqrt{s} = 2 \textrm{TeV}$}}
\put(1068,680){\makebox(0,0)[l]{$\tan\beta = 30$}}
\put(511,663){\makebox(0,0)[l]{$b\bar{b}$}}
\put(511,365){\makebox(0,0)[l]{$gg$}}
\put(201.0,123.0){\rule[-0.200pt]{0.400pt}{157.549pt}}
\put(1279,737){\makebox(0,0)[r]{ }}
\put(201,679){\raisebox{-.8pt}{\makebox(0,0){$\Diamond$}}}
\put(232,669){\raisebox{-.8pt}{\makebox(0,0){$\Diamond$}}}
\put(263,663){\raisebox{-.8pt}{\makebox(0,0){$\Diamond$}}}
\put(294,655){\raisebox{-.8pt}{\makebox(0,0){$\Diamond$}}}
\put(325,647){\raisebox{-.8pt}{\makebox(0,0){$\Diamond$}}}
\put(356,640){\raisebox{-.8pt}{\makebox(0,0){$\Diamond$}}}
\put(387,635){\raisebox{-.8pt}{\makebox(0,0){$\Diamond$}}}
\put(418,629){\raisebox{-.8pt}{\makebox(0,0){$\Diamond$}}}
\put(449,623){\raisebox{-.8pt}{\makebox(0,0){$\Diamond$}}}
\put(480,616){\raisebox{-.8pt}{\makebox(0,0){$\Diamond$}}}
\put(511,606){\raisebox{-.8pt}{\makebox(0,0){$\Diamond$}}}
\put(541,602){\raisebox{-.8pt}{\makebox(0,0){$\Diamond$}}}
\put(572,593){\raisebox{-.8pt}{\makebox(0,0){$\Diamond$}}}
\put(603,588){\raisebox{-.8pt}{\makebox(0,0){$\Diamond$}}}
\put(634,577){\raisebox{-.8pt}{\makebox(0,0){$\Diamond$}}}
\put(665,572){\raisebox{-.8pt}{\makebox(0,0){$\Diamond$}}}
\put(696,564){\raisebox{-.8pt}{\makebox(0,0){$\Diamond$}}}
\put(727,557){\raisebox{-.8pt}{\makebox(0,0){$\Diamond$}}}
\put(758,548){\raisebox{-.8pt}{\makebox(0,0){$\Diamond$}}}
\put(789,545){\raisebox{-.8pt}{\makebox(0,0){$\Diamond$}}}
\put(820,539){\raisebox{-.8pt}{\makebox(0,0){$\Diamond$}}}
\put(851,533){\raisebox{-.8pt}{\makebox(0,0){$\Diamond$}}}
\put(882,528){\raisebox{-.8pt}{\makebox(0,0){$\Diamond$}}}
\put(913,523){\raisebox{-.8pt}{\makebox(0,0){$\Diamond$}}}
\put(944,515){\raisebox{-.8pt}{\makebox(0,0){$\Diamond$}}}
\put(975,509){\raisebox{-.8pt}{\makebox(0,0){$\Diamond$}}}
\put(1006,503){\raisebox{-.8pt}{\makebox(0,0){$\Diamond$}}}
\put(1037,496){\raisebox{-.8pt}{\makebox(0,0){$\Diamond$}}}
\put(1068,488){\raisebox{-.8pt}{\makebox(0,0){$\Diamond$}}}
\put(1099,479){\raisebox{-.8pt}{\makebox(0,0){$\Diamond$}}}
\put(1130,474){\raisebox{-.8pt}{\makebox(0,0){$\Diamond$}}}
\put(1160,469){\raisebox{-.8pt}{\makebox(0,0){$\Diamond$}}}
\put(1191,463){\raisebox{-.8pt}{\makebox(0,0){$\Diamond$}}}
\put(1222,457){\raisebox{-.8pt}{\makebox(0,0){$\Diamond$}}}
\put(1253,453){\raisebox{-.8pt}{\makebox(0,0){$\Diamond$}}}
\put(1284,442){\raisebox{-.8pt}{\makebox(0,0){$\Diamond$}}}
\put(1315,438){\raisebox{-.8pt}{\makebox(0,0){$\Diamond$}}}
\put(1346,434){\raisebox{-.8pt}{\makebox(0,0){$\Diamond$}}}
\put(1377,425){\raisebox{-.8pt}{\makebox(0,0){$\Diamond$}}}
\put(1408,419){\raisebox{-.8pt}{\makebox(0,0){$\Diamond$}}}
\put(1439,414){\raisebox{-.8pt}{\makebox(0,0){$\Diamond$}}}
\put(1349,737){\raisebox{-.8pt}{\makebox(0,0){$\Diamond$}}}
\put(1279,696){\makebox(0,0)[r]{ }}
\put(201,450){\makebox(0,0){$+$}}
\put(232,438){\makebox(0,0){$+$}}
\put(263,425){\makebox(0,0){$+$}}
\put(294,414){\makebox(0,0){$+$}}
\put(325,401){\makebox(0,0){$+$}}
\put(356,385){\makebox(0,0){$+$}}
\put(387,375){\makebox(0,0){$+$}}
\put(418,365){\makebox(0,0){$+$}}
\put(449,352){\makebox(0,0){$+$}}
\put(480,339){\makebox(0,0){$+$}}
\put(511,330){\makebox(0,0){$+$}}
\put(541,315){\makebox(0,0){$+$}}
\put(572,305){\makebox(0,0){$+$}}
\put(603,299){\makebox(0,0){$+$}}
\put(634,287){\makebox(0,0){$+$}}
\put(665,271){\makebox(0,0){$+$}}
\put(696,261){\makebox(0,0){$+$}}
\put(727,271){\makebox(0,0){$+$}}
\put(758,281){\makebox(0,0){$+$}}
\put(789,279){\makebox(0,0){$+$}}
\put(820,275){\makebox(0,0){$+$}}
\put(851,266){\makebox(0,0){$+$}}
\put(882,261){\makebox(0,0){$+$}}
\put(913,253){\makebox(0,0){$+$}}
\put(944,244){\makebox(0,0){$+$}}
\put(975,237){\makebox(0,0){$+$}}
\put(1006,230){\makebox(0,0){$+$}}
\put(1037,221){\makebox(0,0){$+$}}
\put(1068,212){\makebox(0,0){$+$}}
\put(1099,206){\makebox(0,0){$+$}}
\put(1130,201){\makebox(0,0){$+$}}
\put(1160,196){\makebox(0,0){$+$}}
\put(1191,185){\makebox(0,0){$+$}}
\put(1222,179){\makebox(0,0){$+$}}
\put(1253,172){\makebox(0,0){$+$}}
\put(1284,165){\makebox(0,0){$+$}}
\put(1315,156){\makebox(0,0){$+$}}
\put(1346,147){\makebox(0,0){$+$}}
\put(1377,136){\makebox(0,0){$+$}}
\put(1408,130){\makebox(0,0){$+$}}
\put(1439,123){\makebox(0,0){$+$}}
\put(1349,696){\makebox(0,0){$+$}}
\end{picture}
\caption{Total cross section for the process
$p \bar{p} \rightarrow W^{\pm} H^{\mp} + X$ as a function
of $m_H$ via $b\bar{b}$ annihilation and $gg$ fusion
at the Tevatron energy
($\sqrt{s} = 2 \textrm{TeV/c}$) for
$\tan\beta = 30$. Taken from \cite{Barrientos_K}.}
\label{Tevatron_figure}
\end{center}
\end{figure}
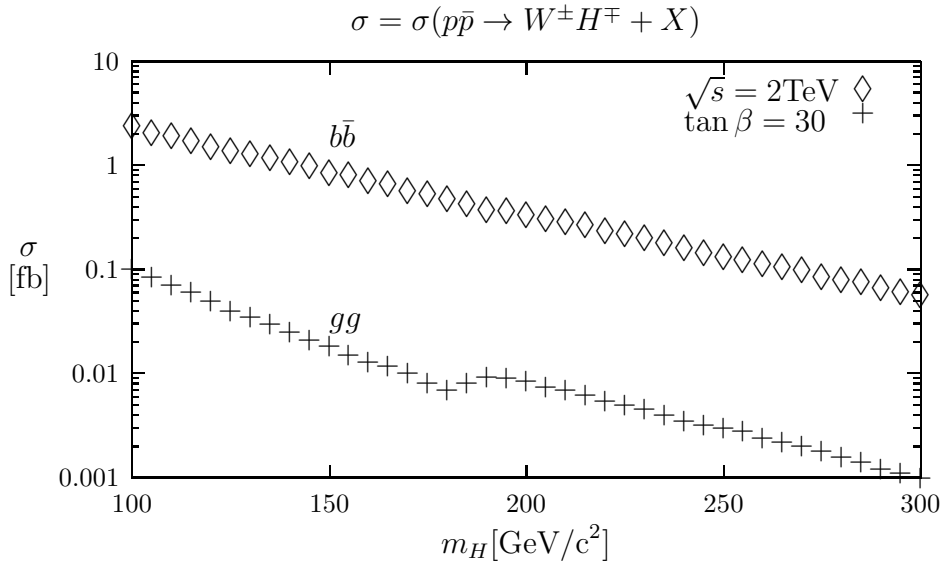
\section{Comparison between $\mu^- \mu^+ \rightarrow H^{\mp} W^{\pm}$
and $ p \bar{p}, pp \rightarrow H^{\mp} W^{\pm}X$ for large values of 
$\tan\beta$}
Let us compare the channel  $\mu^- \mu^+ \rightarrow H^{\mp} W^{\pm}$
at $\sqrt{s} = 500 \textrm{GeV/c}$
with the processes $ p \bar{p}, pp \rightarrow H^{\mp} W^{\pm}X$ at the 
Tevatron energy ($\sqrt{s} = 2 \textrm{TeV/c}$) and LHC energies 
($\sqrt{s} = 14 
\textrm{TeV/c}$) respectively for large values of $\tan\beta$ (for example 
$\tan\beta= 30$).

At the FNAL energy  (Figure \ref{TeVmuon_figure}), we have:
$\sigma(\mu^- \mu^+ \rightarrow H^{\mp} W^{\pm})
> \sigma( p \bar{p} \rightarrow W^{\pm} H^{\mp}X)$ for $\tan\beta= 30$.

At LHC energies (Figure \ref{LHCmuon_figure}), we have:
$\sigma( p p \rightarrow W^{\pm} H^{\mp}X) > \sigma(\mu^- \mu^+ 
\rightarrow H^{\mp}W^{\pm})$ for $\tan\beta=30$.

According to Figure  \ref{muhw_figure}, $\sigma(\mu^- \mu^+
\rightarrow H^{\mp}W^{\pm}) \gtrsim 5 \textrm{fb}$ for $\tan\beta \geq 20$ 
in the 
mass interval $ 100 \leq m_{H} \leq 400 [\textrm{GeV/c}^2]$, which would 
be an observable 
number of $H^{\pm}$ for luminosities $>50 \textrm{fb}^{-1}$. In the mass 
region 
of interest shown in the figures, the dominant decay mode of $H^{\pm}$ is
$H^+ \rightarrow t\bar{b}$ or $H^- \rightarrow \bar{t} b$. So the main 
background would be from $t\bar{t}$ production. Reference 
\cite{S.Moretti} shows that such a background overwhelms the charged
Higgs boson signal in $p \bar{p} \rightarrow W^{\pm} H^{\mp}X$ at the LHC.
In fact, in Section 7 we have shown that $\sigma(\mu^- \mu^+ \rightarrow t 
\bar{t}) \approx 495 \textrm{fb}$ for $\sqrt{s} = 500 \textrm{GeV/c}$. In 
the LHC
the background due to $t \bar{t}$ production is of order \cite{S.Moretti} 
800 pb (three 
orders of magnitude larger than at a muon collider with $\sqrt{s} = 500 
\textrm{GeV/c}$). At the FNAL energy ($\sqrt{s} = 2 \textrm{TeV/c}$) 
something 
similar happens
because $\sigma(p \bar{p} \rightarrow t \bar{t}) = 5.5 \textrm{pb}$ 
\cite{Abazovetal}.

In the muon collider, the signal of the charged Higgs boson is not 
overwhelmed.

Then, for large values of $\tan\beta$, the process $\mu^- \mu^+ 
\rightarrow H^{\mp} W^{\pm}$ is a very attractive channel for the 
search of $H^{\pm}$ at a $\mu^- \mu^+$ collider.
 
\begin{figure}
\begin{center}
\input{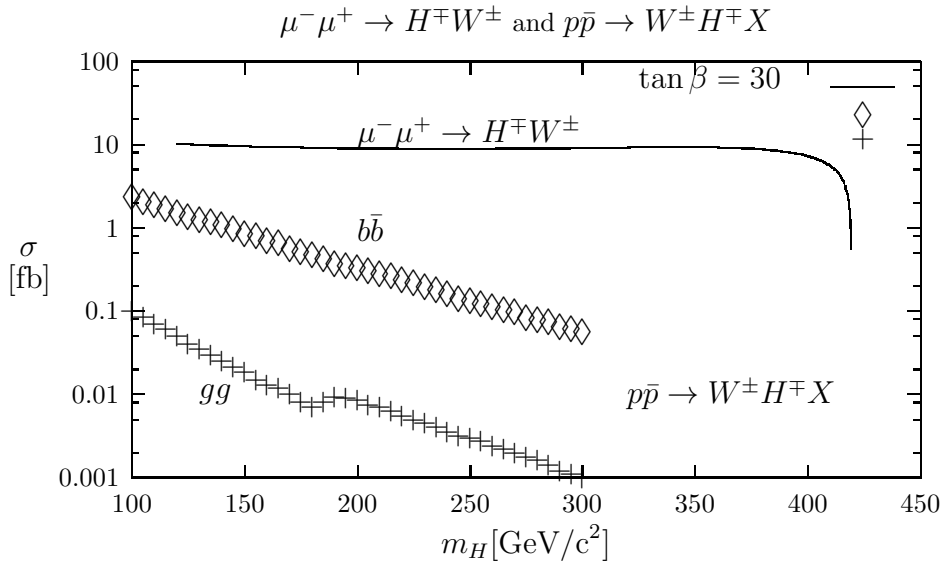}
\caption{Total cross section for the processes
$\mu^- \mu^+ \rightarrow H^{\mp} W^{\pm}$ and
$p \bar{p} \rightarrow W^{\pm} H^{\mp} + X$ (via $b\bar{b}$ annihilation 
and $gg$ fusion) as a function of $m_{H}$
at $\sqrt{s} = 500 \textrm{GeV/c}$ and
$\sqrt{s} = 2 \textrm{TeV/c}$, respectively,  for
$\tan\beta = 30$. Taken partially from \cite{Barrientos_K}.}
\label{TeVmuon_figure}
\end{center}
\end{figure}

\begin{figure}
\begin{center}
\input{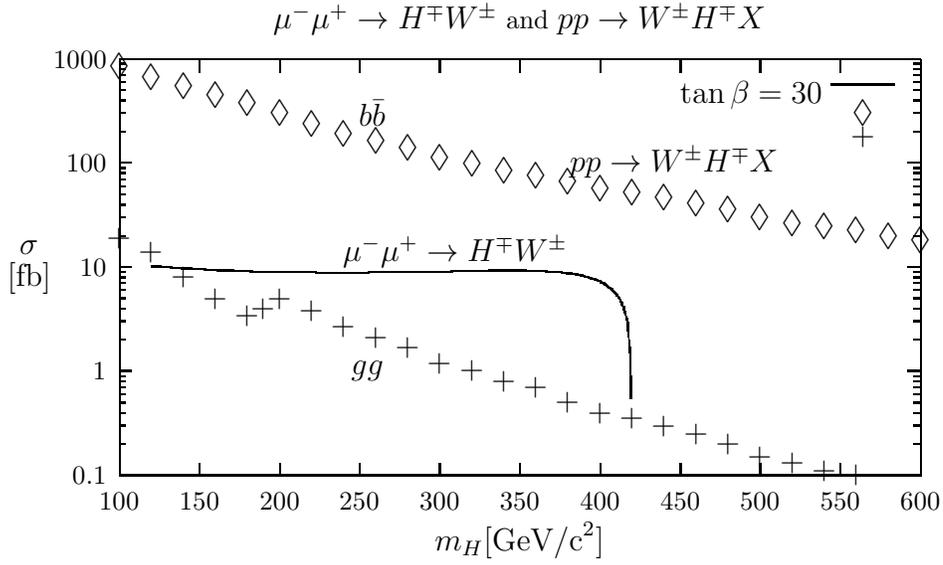}
\caption{Total cross section for the processes
$\mu^- \mu^+ \rightarrow H^{\mp} W^{\pm}$ and
$p p \rightarrow W^{\pm} H^{\mp} + X$ (via $b\bar{b}$ annihilation
and $gg$ fusion) as a function of $m_{H}$
at $\sqrt{s} = 500 \textrm{GeV/c}$ and
$\sqrt{s} = 14 \textrm{TeV/c}$, respectively,  for
$\tan\beta = 30$. Taken partially from \cite{Barrientos_K}.}
\label{LHCmuon_figure}
\end{center}
\end{figure}
\section{Conclusions}
The discovery of the Standard Model Higgs is one of the principal goals
of experimental and theoretical particle physicists. This is because the
Higgs mechanism is a cornerstone of the Standard Model. The search for the
Standard Model Higgs will also constrain or discover particles of the Two
Higgs Doublet Model of type II.

In this paper we have discussed the masses of the Higgs particles in the
Two Higgs Doublet Model of type II, and considered the influence of the 
radiative corrections on these masses. In the absence of radiative 
corrections, the Higgs boson $h^0$ obeys the bound $m_{h^0} \leq m_Z$. 
This bound practically has been excluded by the present limits on 
$m_{h^0}$ obtained by LEP and CDF \cite{LEP_CDF}. However, when the 
radiative corrections are taken into account, $m_{h^0}$ increases as the 
value of $m_{A}$ increases. As a result,  we have a new bound: $m_{h^0} 
\leq 128.062
\textrm{GeV/c}^2$ taking $M_{sb}$ (sbottom mass) and $M_{st}$ (stop mass) 
of order
$1 \textrm{TeV/c}^2$.

Considering the radiative corrections of the masses, we have calculated 

Higgs production 
cross sections at a muon collider in the Two Higgs Doublet Model of type 
II. The most interesting production channels are $\mu^- \mu^+ \rightarrow 
h^0 Z^0, H^0 Z^0$
, $H^- H^+, A^0 Z^0$ and $H^{\mp} W^{\pm}$.  In the first 
two channels the radiative corrections of the masses play an important 
role, which is 
not true for the other channels. In the reaction  $\mu^- \mu^+ \rightarrow
h^0 Z^0$, the total cross section becomes important
in the mass interval $118 \leq m_{h^0} \leq 128 [\textrm{GeV/c}^2]$.
  
The process $ \mu^- \mu^+ \rightarrow A^0 Z^0$, would provide an 
alternative way for searching the $A^0$ looking for peaks in the $b 
\bar{b}$ distribution. Another interesting channel could be $\mu^- \mu^+ 
\rightarrow A^0 h^0$. However, this is highly supressed for $m_{A} \geq 
200 \textrm{GeV/c}^2$ because the total cross section is proportional to 
the factor 
\begin{equation}
\cos^2\left( \beta - \alpha \right) = \frac{\left( 1 + \tan\beta 
\tan\alpha\right)^2}{\left(1+\tan^2\beta\right)\left(1+\tan^2\alpha\right)}
\nonumber
\end{equation}
(see the Feynman rules given in \cite{M_H}). This factor 
decreases as the mass of the $A^0$ increases.

The most attractive channel is $\mu^- \mu^+ \rightarrow H^{\mp} W^{\pm}$, 
see Figures \ref{TeVmuon_figure} and \ref{LHCmuon_figure}. In this 
reaction $\sigma(\mu^- \mu^+ \rightarrow H^{\mp} W^{\pm}) \gtrsim 5 
\textrm{fb}$ 
for
$ \tan\beta \geq 20$ in the mass interval $100 \leq m_{H} \leq 400 
[\textrm{GeV/c}^2]$, which would give an observable number of $H^{\pm}$ 
for 
luminosities $>50 \textrm{fb}^{-1}$ at $\sqrt{s} = 500 \textrm{GeV/c}$.

Because the main background in a hadron collider in the reactions 
$p\bar{p} \rightarrow W^{\pm} H^{\mp}X$ (Tevatron energy) or $ pp 
\rightarrow W^{\pm} 
H^{\mp}X$ (LHC energies) comes from $t \bar{t}$ production, the charged 
Higgs boson 
signal would be overwhelmed by such a background. In a muon collider
with $\sqrt{s} = 500 \textrm{GeV/c}^2$, the signal of the $H^{\pm}$ is not 
overwhelmed. This means, that for large values of $\tan\beta$, the channel
$\mu^- \mu^+ \rightarrow H^{\mp} W^{\pm}$ is a very attractive channel for 
the search of charged Higgs bosons at a $\mu^- \mu^+$ collider.

\section*{Acknowledgment}
I would like to thank Bruce Hoeneisen for the critical reading of this

\noindent manuscript.


\begin{thebibliography} {50}

\bibitem{B-H} Bruce Hoeneisen, Serie de Documentos USFQ $\bf{26}$, 
Universidad San Francisco de Quito, Ecuador (2001).

\bibitem{mu1} J. Gunion, hep-ph/9802258; V. Barger, hep-ph/9803480.

\bibitem{mu2} A. G. Akeroyd, A. Arhrib, C. Dove, Phys. Rev. D $\bf{61}$, 
071702 (2000).

\bibitem{S.Moretti} Stefano Moretti, Kosuke Odagiri, Phys. Rev. D
$\bf{59}$, 055008 (1999).

\bibitem{V.Barger} Vernon Barger and Roger Phillips, Collider Physics 
(Addison Wesley, 1988); S. Dawson, J.F. Gunion, H.E. Haber and G. Kane, 
The Higgs Hunter's Guide (Addison Wesley, 1990).

\bibitem{LEP_CDF} Review of Particle Physics, K. Hagiwara et al., Phys. 
Rev. D $\bf{66}$, 010001 (2002).

\bibitem{Weinberg} "The quantum theory of fields", Volume III, 
Supersymmetry, Steven Weinberg, Cambridge University Press (2000).

\bibitem{Zhou} Zhou Fei et al., Phys. Rev. D $\bf{64}$, 055005(2001).

\bibitem{M_H} C. Mar\'{\i}n and B. Hoeneisen, 
hep-ph/0402061 v1 (2004).

\bibitem{CM} C. Mar\'{\i}n, Polit\'{e}cnica, $\bf{XVII}$ No. 1, p.79 
(1992), Escuela Polit\'{e}cnica Nacional, Quito, Ecuador. DO Note 
Fermilab (1992).

\bibitem{Barrientos_K} A. A. Barrientos Bendez\'{u} and B. A. Kniehl, 
Phys. Rev. D $\bf{59}$, 015009 (1998).

\bibitem{Abazovetal} V. M. Abazov et al., Phys. Rev. Lett. $\bf{88}$, 
151803 (2002).

\end{thebibliography}
\end{document}